\documentclass[12pt]{amsart}
\usepackage{amsmath}
\usepackage{amsxtra}
\usepackage{amscd}
\usepackage{amsthm}
\usepackage{amsfonts}
\usepackage{amssymb}
\usepackage{eucal}
\usepackage{epsfig}
\usepackage{graphics}
\usepackage[dvips]{color}
\usepackage{epsf,graphics,mathrsfs,yfonts,eufrak,simplewick}
\usepackage{pdfpages}
\usepackage{psfrag}
\usepackage[OT2,T1]{fontenc}

\def\'{\char126}
\def\`{\char127}

\def\({\left(}
\def\){\right)}
\newcommand{\mub}{\mbox{\boldmath$\mu$}}

\newcommand{\taub}{\mbox{\boldmath$\tau$}}

\newcommand{\sgn}{\mathop{\rm sgn}}

\newcommand{\nn}{\nonumber}

\newcommand{\slth}{\widehat{\mathfrak{sl}}_2}
\newcommand{\res}{{\rm res}}

\newenvironment{tenumerate}{
  \begin{enumerate}
  
  }{\end{enumerate}}
\newcommand{\bi}{\begin{tenumerate}}
\newcommand{\ei}{\end{tenumerate}}
\newcommand{\isoto}[1][]%
{{\mathop{\buildrel{\sim}\over\longrightarrow}\limits_{#1}}}

\def\[{\left[}
\def\]{\right]}

\newcommand{\al}{\alpha}

\newcommand{\z}{\zeta}

\numberwithin{equation}{section}

\def\half{\textstyle{\frac  1 2}}

\def\bi{\mathbf{i}}

\newcommand{\betab}{\mbox{\boldmath$\beta$}}
\newcommand{\gammab}{\mbox{\boldmath$\gamma$}}

\newcommand{\Rho}{\varrho}

\usepackage[dvipsnames]{xcolor}

\begin{document}
\begin{title}[New results on integrable structure]
{New results on   integrable structure of conformal field theory
}
\end{title}
\date{\today}
\author{H.~Boos and  F.~Smirnov}
\address{HB: Physics Department, University of Wuppertal, D-42097,
Wuppertal, Germany}\email{boos@physik.uni-wuppertal.de}
\address{FS\footnote{Membre du CNRS}: 
${}^{1}$ Sorbonne Universit\'e, UPMC Univ Paris 06\\ CNRS, UMR 7589, LPTHE\\F-75005, Paris, France}\email{smirnov@lpthe.jussieu.fr}

\dedicatory{In memory of Petr Petrovich Kulish}
\begin{abstract}

We explain how to incorporate the action of local integrals of motion into the fermionic
basis for the sine-Gordon model and its UV CFT. The examples up to the level 4 are presented.
Numerical computation support the results. Possible applications are discussed.
\end{abstract}

\maketitle

\section{Introduction}

The fermionic basis in the space of local operators appeared in the papers \cite{HGS,HGSII,HGSIII}
in the study of XXZ spin chain. The important feature of this basis is that the expectation
values on a cylinder are given by determinants. Later it was realised \cite{HGSIV,HGSV} that
the fermionic basis provides a universal description of integrable models related to
$U_q(\slth)$, namely XXZ (homogeneous or inhomogeneous) lattice model,
CFT with $c<1$, sine-Gordon (sG) model. In certain sense the fermionic basis provides
an invariant under the action of the renormalisation group description of these integrable models.

It is especially important to find exact quantitative relation between the fermionic basis
and the usual Virasoro description of the CFT. This is due to fact that the study of 
non-trivial asymptotics of the
correlation functions (short-distance asymptotics for massive relativistic models
and long-distance asymptotics for lattice models) passes through the perturbed CFT (PCFT)
\cite{Alyosha2p,FFLZZ,Lukyanov}. 

In the papers \cite{OP,HGSV} it was shown that the fermionic basis allows to compute the
one-point functions of descendants for the sG model in infinite volume and on the 
infinite cylinder $\mathbb{C}/2\pi i R  \mathbb{Z}$
(equivalently at finite temperature $T=1/(2\pi R)$) providing an indispensable set of data for study of the 
UV asymptotics of
correlation functions. Conventionally  the directrix (generatrix) of  the cylinder is called  Matsubara (Space) direction.
The potential in the sG Lagrangian is periodic, hence there is the  Floquet index $p$.
We treat the Euclidean sG model \cite{HGSV} as the scaling limit of the inhomogeneous
six-vertex model. This approach is technically 
closer to the dual massive Thirring model (MTM) than to the sG model. For the
MTM introducing the Floquet index corresponds to putting a finite
gauge transformation  into the boundary conditions in Matsubara direction
for the Dirac fermion:
$\psi(x+2\pi i R)=e^{2\pi ip}\psi(x)$. Alternatively, we can introduce a constant gauge field along
the Matsubara direction.
We shall call $p$ twist. 
Considering the sG model \cite{HGSV}  on a cylinder it was  implied that the values of
$p$ coincide for $\mathrm{Re}{z}\to\pm \infty$.
In this case it is sufficient to know the identification of the fermionic basis with the
Virasoro one modulo action of the local integrals of motion. 
However, there are situations when this is not sufficient.

Consider  a cylinder  and  impose
different boundary conditions: different twists $p$ and $p'$ for 
$\mathrm{Re}{z}\to \infty$ ans $\mathrm{Re}{z}\to -\infty$. 
Certainly, the ``partition function" with such boundary conditions vanish due 
to orthogonality of eigenstates with different twists. 
However, the insertion of a local operators $\mathcal{O}(0)$ makes the result non-trivial. 
The resulting one-point function is the closet relative of three-point function in the UV CFT
which should be reproduced in the high temperature limit. This is what we shall observe
numerically in Section \ref{numerics}.

It is also interesting to identify on quantitive level
the form factors with the all the local operators in the theory. 
The form factors by definition are computed between different states, hence
the action of local operators in non-trivial.
The form factors in the fermionic basis were described in \cite{JMSFF}, so,
knowing the exact relation between the fermionic basis and the Virasoro
descendants we can find the form factors of the latter ones.

Finally, the knowledge
of complete (not modulo integrals of motion) fermionic basis in CFT should be
useful for obtaining a regular way of computing the long-distance asymptotics
for the lattice XXZ model through PCFT in the spirit of \cite{Lukyanov}. 
We hope to return to the latter problem in future. 

Returning to  the computations modulo action of integrals of motion, 
there were two ways to proceed. The expectation values of local operators 
in fermionic description are given by determinants containing one function of
two variables $\omega(\z,\xi)$. 
Relying on the ideas of \cite{BLZII} a recursive procedure for computing of asymptotics
of this
function was  developed in \cite{HGSIV}. This allowed to identify the fermionic basis with
the  Virasoro one up to the level 6. Later this was pushed to the level 8 \cite{Boos}. 

Another way of doing is based on usage of the reflection relations \cite{FFLZZ}. 
These relations were conjectured for the computation of one-point functions.
The problem with them is that they lead to a complicated Riemann-Hilbert problem, and for some time
it was not clear how to solve it. In \cite{NS} it was shown that the known formulae for
the fermionic basis provide the solution up to the level 8. Moreover, combining the
Riemann -Hilbert problem with the conjectured existence of the fermionic basis
allows to proceed further: the fermionic basis on level 10 was obtained as a demonstration
of power of the method. 

Immediately after appearance of \cite{NS} an idea occurred to us to apply the
reflection relations in order to remove a limitation of working modulo
action of the integrals of motion. Indeed, known results together with the
reflection relations lead to some linear difference equations for the unknown
coefficients (those in front of descendants created by the local integrals). 
Assuming analyticity one easily solve these equations. However, the problem
is in possibility of adding quasi-constants. We were unable to fix them
in order to agree with the known description of null-vectors \cite{JMSFF}.

This was very discouraging, and we were looking for some additional information
from the study of the function $\omega(\z,\xi)$ for the case of different
boundary conditions at $\mathrm{Re}(z)\to\pm\infty$ on a cylinder.
It seems hardly possible to apply directly for CFT a procedure similar to that of \cite{HGSIV}
to study the asymptotics of $\omega(\z,\xi)$ (at $\xi,\z\to 0$ or $\infty$) in this case for certain
technical problems. On the other hand one can try to compute numerically this function for the 
sG case and to consider the high temperature limit
making contact with CFT. This is hard, but doable
as we shall explain in Section \ref{numerics}. Comparing the numerical results with the
conjectured reflection relations we found correct formulae for the fermionic
basis up to the level 4. Then we realised what was going wrong in our comparison
with the null-vectors. 

In the present paper we report our results following not this 
``inductive" way, but rather ``deductive" one. We correct the definition
of the fermionic basis in Section \ref{first} , and conjecture its relation with the Virasoro one
up to the level 4 in Sections \ref{l2},\ref{l4}. In the same sections we demonstrate agreement with the null-vectors.
We start to discuss the function $\omega(\z,\xi)$ in the case of different boundary conditions
and explain how to relate its asymptotics to the expectation values in Section \ref{expectation}.
Then in the Section \ref{numerics} we show that our conjectures are in perfect agreement
with the numerical data.

\section{Some formulae about form factors and null-vectors}\label{first}

We consider the Euclidean sine-Gordon (sG) model as perturbation of ``complex Liouville model" (cL):
\begin{align}
\mathcal{A}^\mathrm{sG}&=\int \left\{ \Bigl[\frac 1 {4  \pi} \partial _z\varphi (z,\bar{z})\partial_{\bar{z}}\varphi (z,\bar{z})-
 \Lambda e^{-i\beta\varphi(z,\bar{z})}\Bigr]
-
\Lambda   e^{i\beta\varphi(z,\bar{z})}    \right\}d^2 z\,.
\label{action1}
\end{align}
As explained in \cite{HGSV} it is convenient to use as dimensional parameter $\mub$ related to $\Lambda$ by
$$\Lambda=\frac{\mub ^2}{\sin\pi\beta ^2}\,,$$
We introduce the  XXZ-like parameter
$$\nu =1-\beta^2\,,$$
as coupling constant. 
In this paper we shall assume $1/2<\nu<1$, but everything can be done for $0<\nu<1/2$ as well.
Notice that $\mub^{\frac 1 \nu}$ has dimension of mass.

For generic $\al$ the local operators in sG are identified with their
cL counterparts unambiguously. So, we do not distinguish them notationally. 
In the cL Liouville theory we have primary fields and their descendants. It will be convenient to
work with dimensionless operators. The primary fields are
$$\Phi_\al(0)=\mub^{-\frac {\nu}{2(1-\nu)}\al^2}e^{\frac \nu{2(1-\nu)}\al\(i\beta\varphi(0)\)}\,,$$
the descendants are created by 
$\mathbf{l}_{-k}$ defined form the energy-momentum tensor via
$$
T(z)=\mub^{\frac 2 \nu}\sum_{n=-\infty}^{\infty}(\mub^{\frac 1 \nu} z)^{-n-2}\mathbf{l}_n\,.
$$

For economy of Greek letters form now on we use  for the coupling constant only $\nu$. 

Combine our fermions and integrals of motion into the generating functions
\begin{align}
\betab^*(\z)=\sum_{j=1}^\infty\betab^*_{2j-1}\z^{-\frac{2j-1}\nu}\,,\quad
\gammab^*(\z)=\sum_{j=1}^\infty\gammab^*_{2j-1}\z^{-\frac{2j-1}\nu}\,,\quad \taub^*(\z)=\exp\(\sum_{j=1}^\infty\taub^*_{2j-1}\z^{-\frac{2j-1}\nu}\)\,,
\end{align}
where
\begin{align}\taub^*_{2j-1}=&C_{2j-1}(\nu)\mathbf{i}_{2j-1}\,,\label{taub}\\&
C_{2n-1}(\nu)=-s(\nu)^{2n-1}\frac{\sqrt{\pi(1-\nu)}}{\nu n!}\frac{\Gamma\(\frac{2n-1}{2\nu}\)}{\Gamma\(1+\frac{(2n-1)(1-\nu)}{2\nu}\)}\nn\,,
\end{align}
here and later
$$s(\nu)=\Gamma(\nu)^{-\frac 1\nu}\sqrt{1-\nu}\,.$$
The operators $\taub^*_{2j-1}$ belong to the centre, so, we use them as constants.

We introduce formally the annihilation operators 
\begin{align}
\betab(\z)=\sum_{j=1}^\infty\betab_{2j-1}\z^{\frac{2j-1}\nu}\,,\quad
\gammab(\z)=\sum_{j=1}^\infty\gammab_{2j-1}\z^{\frac{2j-1}\nu}\,.\nn
\end{align}
Similar operators are introduced for the second chirality with the change $\z\to1/\z$.

These operators, as they has been defined in \cite{HGSIV}, do not satisfy the reflection relations, they have to be slightly modified. Namely to any of our operators we apply the similarity transformation defining
\begin{align}
\mathbf{x}_\mathrm{new}(\z)=e^{\Omega_{\mathrm{cor}}}\mathbf{x}_\mathrm{old}(\z)e^{-\Omega_{\mathrm{cor}}}\,,
\label{change}
\end{align}
with
\begin{align}
\Omega_{\mathrm{corr}}=\frac i{2 \nu}\cot\({\textstyle \frac \pi 2} \al\)
\Bigl(\mathcal{Q}+\overline{\mathcal{Q}}\Bigr)\Bigl(\mathcal{Q}^\dag+\overline{\mathcal{Q}}^\dag\Bigr)
\,,
\end{align}
where
$$
\mathcal{Q}=\frac 1 {(2\pi i)}\oint(\taub^*(\z))^{\frac 1 2}\betab(\z)
\frac{d \z^{\frac 1 \nu}}{ \z^{\frac 1 \nu}}\,,\quad
\mathcal{Q}^\dag=\frac 1 {(2\pi i)}\oint(\taub^*(\z))^{\frac 1 2}\gammab(\z)
\frac{d \z^{\frac 1 \nu}}{ \z^{\frac 1 \nu}}\,,
$$
and similarly for other chirality. 
This modification is irrelevant in   \cite{HGSIV,HGSV,NS} because in these papers all
the serious computation were done for the case when $\taub^*(\z)$ is essentially equal to $1$ in which case
$\mathcal{Q}=0$ {\it etc}. However, the change \eqref{change} can be felt in the context of \cite{JMSFF} where the
form factors are considered. Let us present modified formulae. 

We recall some formulae from \cite{JMSFF}. Form factors are given by certain integral transformation (see \cite{book,JMSFF} for details). Different descendants of
the primary field $\Phi_\al$
are labeled by
certain anti-symmetric functions inserted into the integral. Let us describe them.
Consider the state of $2n$ solitons with rapidities $\beta_1,\cdots\beta_{2l}$, and
introduce 
$$P(S)=\prod_{j=1}^l{(S-B_j)}\,,\quad B_j=e^{\beta_j}$$
Here and later
$$\z=Z^{\nu}\,\quad \xi =X^{\nu},\ \ etc.$$
The central operator $\taub^*(\z)$ is evaluated in this case as $P(Z)/P(-Z)$.
The above mentioned functions  are constructed using 
\begin{align*}
&C_{\pm}(S_1,S_2)=
\frac 1 2\sum_{\epsilon_1,\epsilon_2=\pm}
\epsilon_1\epsilon_2
P(\epsilon_1 S_1)P(\epsilon_2 S_2)
\tau_\pm(\epsilon_2 S_2/\epsilon_1 S_1,\al)\,,
\end{align*}
where
\begin{align}
&\tau_+(x,\al)=-\Bigl(       \half  t_0(\al)+ \sum\limits_{p=1}^{\infty}(-x)^p t_p(\al)\Bigr)
\,,\quad
&\tau_-(x,\al)=\half  t_0(\al)+\sum\limits^{-1}_{p=-\infty}(-x)^p
t_p(\al)\,,
\label{taus}
\end{align}
here and later
\begin{align}
t_p(\al)={\textstyle\frac i {2\nu}}
\cot{\textstyle\frac \pi 2}(\al +{\textstyle\frac p\nu})\,.\label{deft}
\end{align}

The form factors of the operator
\begin{align}
\mathcal{O}_\al=\betab^*(\z_1)\cdots \betab^*(\z_p)\bar{\betab}^*(\z_{p+1})\cdots\bar{ \betab}^*(\z_k)
\bar{\gammab}^*_{k'}(\xi_{k'})\cdots\bar{\gammab}^*(\xi_{q+1})\gammab^*(\xi_q)\cdots\gammab^*_1(\xi_1)\Phi_\al\,,\nn
\end{align}
are associated with the anti-symmetric functions
\begin{align}
L_{O_\al}(S_1,\cdots,S_l)&=\langle\Phi _\al\rangle
\frac{1}{\prod\limits_{j=1}^k\sqrt{P(Z_j)P(-Z_j)}\ \ 
\prod\limits_{j=1}^{k'}\sqrt{P(X_j)P(-X_j)}}\label{polyn}
\\&\times\prod_{j=1}^lS_j^{\frac\nu{1-\nu}\al}\prod_{j=1}^{2l}B_j^{- \frac{\nu}{2(1-\nu)}\al}\cdot\left|\ 
\begin{matrix}
\mathcal{A}&\mathcal{B}\\ \mathcal{C}&\mathcal{D}
\end{matrix}
\ \right|\,,
\end{align}
where $\langle\Phi _\al\rangle$ is the one-point function of the primary field in infinite volume \cite{LZ},
and $\mathcal{A}$, $\mathcal{B}$, $\mathcal{C}$ and $\mathcal{D}$
are respectively 
$k\times k'$, $k\times l$, $n\times k'$ and $n\times l$ 
matrices:
\begin{align}
\mathcal{A}=
\begin{pmatrix}
0&\cdots &0&C _+(Z_1,X_{q+1})&\cdots&C _+(Z_1,X_{k'})\\
\vdots&\   &\vdots &\vdots&\ &\vdots \\
0&\cdots &0&C _+(Z_p,X_{q+1})&\cdots&C _+(Z_p,X_{k'})\\
C _-(Z_{p+1},X_1)&\cdots&C _-(Z_{p+1},X_{q})&0&\cdots &0\\
\vdots&\   &\vdots &\vdots&\ &\vdots \\
C _-(Z_{k},X_1)&\cdots&C _-(Z_{k},X_{q})&0&\cdots &0\\
\end{pmatrix}\,.\nn
\end{align}
\begin{align}
\mathcal{B}=\begin{pmatrix}
C _+(Z_1,S_1)&\cdots &C _+(Z_1,S_l)
\\
\vdots&\   &\vdots \\
C _+(Z_p,S_1)&\cdots &C _+(Z_p,S_l)\\
C _-(Z_{p+1},S_1)&\cdots&C _-(Z_{p+1},S_l)\\
\vdots&\   &\vdots \\
C _-(Z_{k},S_1)&\cdots&C _-(Z_{k},S_l)
\end{pmatrix}\,,\nn
\end{align}
\begin{align}
\mathcal{C}=\begin{pmatrix}
X_{1} &\cdots&X_{k'}\\ 
\vdots &\ &\vdots\\
X^{2l-1}_{1} &\cdots &X^{2l-1}_{k'} 
\end{pmatrix}\,,
\quad
\mathcal{D}=\begin{pmatrix}
S_1 &\cdots&S_l\\
\vdots &\ &\vdots\\
S^{2l-1}_1 &\cdots &S^{2l-1}_l 
\end{pmatrix}\,.\nn
\end{align}
The necessary explanations about this construction is given in \cite{JMSFF}.
The formulae \eqref{taus} are symmetric for two chiralities, contrary to corresponding formulae from \cite{JMSFF}.
This is due to the modification \eqref{change}. This looks as rather innocent modification, but it took for us some effort to
realise its necessity. 

When $\al=m\frac{1-\nu}\nu$ the function \eqref{polyn} becomes a polynomial.
For certain polynomials the integral for the form factors vanish due to $q$-exact one-forms
and $q$-Riemann bilinear identity which lead together to a procedure of describing the null-vectors
in terms of form factors for the primary fields
$\Phi_{1,j}=e^{i(j-1)\beta\varphi/2}$. 
This is explained  in  \cite{BBS,JMSFF}, so, we shall not go into details providing just the
final results.  The construction differs for odd and even $j$. 

{We are dealing with the integrable structure of CFT which is different from the conformal one.
In the conformal description we have singular vectors and
null-vectors which are their Virasoro descendants. In the integrable description the descendants
are created by the action of integrals of motion only. So, the Virasoro null-vectors which
are not obtained in this way have to be described in the fermionic basis. This can be done in
rather compact way explained below for the descendants of $\Phi_{1,2m+1}$  and $\Phi_{1,2m}$.
}

 For $\Phi_{1,2m+1}$ we have $\al =2m\frac{1-\nu}{\nu}$. Introduce two operators:
 \begin{align}
 &\mathcal{Q}_{2m+1}=\frac 1 {2\pi i}\oint \z^{-\frac{2m}{\nu}}\taub^*(\z)^{1/2}\gammab(\z)\cdot\frac{d\z^{\frac 1 \nu}}{\z^{\frac 1 \nu}}\,,\nn\\
 & \mathcal{C}_{2m+1}=\frac 1 {2\pi i}\oint \z^{-\frac{4m}{\nu}}\betab^*(\z)\gammab(\z)\cdot\frac{d\z^{\frac 1 \nu}}{\z^{\frac 1 \nu}}\nn\\&+\frac 1 {(2\pi i)^2}\oint \oint 
 \sum_{\epsilon_j=\pm}\epsilon_1\epsilon_2\taub^*(\z_1)^{\epsilon_1/2}\taub^*(\z_2)^{\epsilon_2/2}\tau_{2m+1}(\epsilon_1\z_1^{\frac 1 \nu},\epsilon_2\z_2^{\frac 1 \nu})\gammab(\z_1)\gammab(\z_2)
\cdot\frac{d\z_1^{\frac 1 \nu}}{\z_1^{\frac 1 \nu}}\cdot\frac{d\z_2^{\frac 1 \nu}}{\z_2^{\frac 1 \nu}}\,,\nn
 \end{align}
were
$$\tau_{2m+1}(X,Y)=-\half\Bigl(
\sum\limits_{{j=1}}^{2m-1}{(-1)^j}t_j(0)X^{-(2m+j)}Y^{-(2m-j)}+\half X^{-4m}t_{2m}(0)\Bigr)\,.$$
Then the null-vectors are of two types:
\begin{align}
&\mathcal{Q}_{2m+1}\(\mathcal{C}_{2m+1}\)^m|\Psi\rangle\,,\quad \(\mathcal{C}_{2m+1}\)^{m+1}|\Psi\rangle\,,\,,
\end{align}
for any descendent $|\Psi\rangle$ of $\Phi_{1,2m+1}$. The null-vector of lowest dimension (singular vector) is
\begin{align}
\mathcal{Q}_{2m+1}\mathcal{C}_{2m+1}^m\gammab^*_{4m+1}\gammab^*_{4m-1}\cdots\gammab^*_1\Phi_{1,2m+1}\,,
\label{singodd}
\end{align}

For $\Phi_{1,2m}$ we have $\al=(2m-1)\frac{1-\nu}\nu$. In this case we need only one operator:
\begin{align}
  & \mathcal{C}_{2m}=\frac 1 {2\pi i}\oint \z^{-\frac{4m-2}{\nu}}\betab^*(\z)\gammab(\z)\cdot\frac{d\z^{\frac 1 \nu}}{\z^{\frac 1 \nu}}\nn\\&+\frac 1 {(2\pi i)^2}\oint \oint 
 \sum_{\epsilon_j=\pm}\epsilon_1\epsilon_2
 \taub^*(\z_1)^{\epsilon_1/2}\taub^*(\z_2)^{\epsilon_2/2}\tau_{2m-2}(\epsilon_1\z_1^{\frac 1 \nu},\epsilon_2\z_2^{\frac 1 \nu})\gammab(\z_1)\gammab(\z_2)
\cdot\frac{d\z_1^{\frac 1 \nu}}{\z_1^{\frac 1 \nu}}\cdot\frac{d\z_2^{\frac 1 \nu}}{\z_2^{\frac 3 \nu}}\,,\nn
\end{align}
were
$$\tau_{2m}(X,Y)=-\half\Bigl(\sum\limits_{j=0}^{2m-1}{(-1)^j}t_j({\textstyle\frac{1-\nu}{\nu}})X^{-(2m+j)}Y^{-(2m-j)}+\half X^{-4m}t_{2m}({\textstyle\frac{1-\nu}{\nu}})\Bigr)\,.$$
Here there is only one kind of null-vectors:
\begin{align}
\( \mathcal{C}_{2m}\)^m|\Psi\rangle\,.\nn
\end{align}
The singular vector is given by
\begin{align}
\( \mathcal{C}_{2m}\)^m\gammab^*_{4m-1}\gammab^*_{4m-1}\cdots\gammab^*_1\Phi_{1,2m}\,,
\label{singodd}
\end{align}

Let us write down  null-vectors up to level 5. Applying the above formulae we obtain results with
several fictitious poles in $\nu$, which is not very convenient for numerical checks;
by some exercises in trigonometry we eliminate them. One more remark concerns the case $\Phi_{1,1}$ ($\al=0$).
At $\al\to 0$ there is a singularity in our formulae (see, for example \eqref{taus}). We can redefine the operators $\betab^*$, $\gammab^* $ in order to eliminate this singularity
{(undoing the transformation \eqref{change} )}, but it is easier to observe that in the 
formulae below singular term always
comes accompanied by pure descendants of integrals of motion which vanish on the unit operator. So, we shall not
introduce modified operators, simply dropping the singular terms.

{Below we collect the results. It has been explained how the singular vectors are obtained, for other null-vectors we explain their origins.}

{On level 1 there is one singular vector 
\begin{align}
\Psi_{1,1}&=\taub^*_1\Phi_{1,1}\,,\label{sing11}
\end{align}
obtained as $\mathcal{Q}_1\gammab^*_1\Phi_{1,1}$. For $\Phi_{1,1}$ there are 
other null-vectors created by the integrals of motion which we shall neglect for their triviality.}

On level 2 we have one singular vector:
\begin{align}
\Psi_{1,2}&=\[\betab^*_1\gammab^*_1-\frac i {8\nu}
\cot {\textstyle\frac{\pi}2 \(\frac{1-\nu}\nu\)}(\taub^*_1)^2\]\Phi_{1,2}\,,\label{sing12}
\end{align}

On level 3 there is one singular vector
\begin{align}
\Psi_{1,3}&=\[\cot {\textstyle\frac{\pi}2 \(\frac{1-\nu}\nu\)}\taub^*_1\betab^*_1\gammab^*_1+\frac i {\nu}
\taub^*_3
-\frac i {48\nu }\(1+ 3\cot {\textstyle\frac{\pi}2 \(\frac{1-\nu}\nu\)}^2\)(\taub^*_1)^3\]\Phi_{1,3}\,,\label{sing13}
\end{align}

On level 4 we have two singular vectors
\begin{align}
\Psi_{1,4}&=\[\betab^*_1\gammab^*_3-\frac i {8\nu}\(\cot {\textstyle\frac{3\pi}2 \(\frac{1-\nu}\nu\)}
+2\cot {\textstyle\frac{\pi}2 \(\frac{1-\nu}\nu\)}\)\taub^*_1\taub^*_3\right.\label{sing14}
\\&\left.-\frac 1 8\cot {\textstyle\frac{\pi}2 \(\frac{1-\nu}\nu\)}^2(\taub^*_1)^2\betab^*_1\gammab^*_1+
\frac i {192\nu}\(\cot {\textstyle\frac{\pi}2 \(\frac{1-\nu}\nu\)}+\cot {\textstyle\frac{\pi}2 \(\frac{1-\nu}\nu\)}^3\)(\taub^*_1)^4\]\Phi_{1,4}\,.\nn
\end{align}
\begin{align}
\Psi_{2,2}&=\[\betab^*_1\gammab^*_3-\betab^*_3\gammab^*_1\]\Phi_{2,2}\,.\label{sing22}
\end{align}
The latter case does not quite fit our previous consideration because we considered the fields $\Phi_{1,k}$ only,
however it can be checked with formulae which we shall present soon. Actually, it is not very hard to guess
how our fermionic description applied to null vectors of all the operators
from the Kac table, we hope to return to this question in another publication. 
There are additional null-vectors (they coincide with certain Virasoro descendants of singular vectors):
\begin{align}
&\[\betab^*_3\gammab^*_1-\frac i{8 \nu}\cot {\textstyle\frac{\pi}2 \(\frac{1-\nu}\nu\)}\taub^*_1\taub^*_3-\frac i {192\nu}\cot {\textstyle\frac{\pi}2 \(\frac{1-\nu}\nu\)}(\taub^*_1)^4\]\Phi_{1,2}\,,\label{null12}
\end{align}
{proportional to $\mathcal{C}_2\gammab^*_5\gammab ^*_1\Phi_{1,2}$.}
\begin{align}
(\betab^*_1\gammab^*_3-\betab^*_3\gammab^*_1)\Phi_{1,1}\,,\label{null11}
\end{align}
{proportional to $\mathcal{C}_1\gammab^*_3\gammab ^*_1\Phi_{1,1}$.}

On level 5 we have the singular vector
\begin{align}
\Psi_{1,5}&=\[\cot {\textstyle\frac{\pi}2 \(\frac{1-\nu}\nu\)}\taub^*_1\betab^*_1\gammab^*_3+\cot {\textstyle\frac{3\pi}2 \(\frac{1-\nu}\nu\)}\,
\taub^*_3\,\betab^*_1\gammab^*_1+\frac{i}\nu\,\taub^*_5-\frac 1{24}\cot {\textstyle\frac{\pi}2 \(\frac{1-\nu}\nu\)}^3(\taub^*_1)^3\betab^*_1\gammab^*_1\right.\label{sing15}\\&-\frac {i}{8\nu}\cot {\textstyle\frac{\pi}2 \(\frac{1-\nu}\nu\)}\(\cot {\textstyle\frac{\pi}2 \(\frac{1-\nu}\nu\)}+\cot {\textstyle\frac{3\pi}2 \(\frac{1-\nu}\nu\)}\)(\taub^*_1)^2\taub^*_3\nn\\&\left.+\frac i{3840 \nu}
\(-3 + 10 \cot {\textstyle\frac{\pi}2 \(\frac{1-\nu}\nu\)}^2 + 
   5 \cot {\textstyle\frac{\pi}2 \(\frac{1-\nu}\nu\)}^4\) ( \taub^*_1)^5\]\Phi_{1,5}\nn\,,
\end{align}
and null-vectors
{\begin{align}
\Bigl(  \taub^*_1\betab^*_1\gammab^*_3-\Bigl(   \taub^*_3+{\textstyle \frac 1 {24}}
 \(   \taub^*_1\)^3\Bigr)\betab^*_1\gammab^*_1\Bigr)\Phi_{1,1}\,,\label{null115}
\end{align}
proportional to $\mathcal{Q}_1\betab^*_1\gammab^*_3\gammab ^*_1\Phi_{1,1}$,}
\begin{align}
&\Bigr[\cot {\textstyle\frac{\pi}2 \(\frac{1-\nu}\nu\)}\taub^*_1\betab^*_3\gammab^*_1
+\frac i {\nu}\taub^*_5-
\frac i {16\nu}(1+\cot {\textstyle\frac{\pi}2 \(\frac{1-\nu}\nu\)}^2)
\taub^*_3(\taub^*_1)^2\label{null13}
\\&+\frac i {1920\nu}(1-5\cot {\textstyle\frac{\pi}2 \(\frac{1-\nu}\nu\)}^2)(\taub^*_1)^5
\Bigl]\Phi_{1,3}\nn\,,
\end{align}
{proportional to $\mathcal{Q}_3\mathcal{C}_3\gammab^*_7\gammab^*_3\gammab ^*_1\Phi_{1,3}$.}

Important property of our fermions consists in the fact that they create not the tensor
product of two chiral Verma modules, but rather the direct sum \cite{HGSV}
$$\bigoplus\limits_{m=-\infty}^{\infty}\mathcal{V}_{\al+2m\frac{1-\nu}{\nu}}\otimes
\overline{\mathcal{V}}_{\al+2m\frac{1-\nu}{\nu}}\,.$$
This property has very transparent explanation in terms of the functions \eqref{polyn}
due to the simple identity
\begin{align}
\tau_+(x,\al)=-\half t_0(\al)+ t_1(\al)x-\half t_2(\al)x^2+x^2\tau_+(x,\al+2{\textstyle\frac{1-\nu}{\nu}})\,.\nn
\end{align}
We do not go into details at this
point because very careful explanation can be found in \cite{JMSFF}, but rather give one example which we shall need later.

For shift of primary field one has 
\begin{align}
&C(\al,\nu)\cdot\betab^*_1\bar{\gammab}^*_1
\Phi_{\al}=2t_1(\al)\cdot \Phi_{\al+2\frac{1-\nu}{\nu}}\,,\label{shift1}
\end{align}
where
\begin{align}
&C(\al,\nu)=
\Gamma(\nu)^{4x}
\frac{\Gamma(-2\nu x)\Gamma( x)\Gamma(\frac1 2 -x)}{\Gamma(2\nu x)\Gamma( -x)\Gamma(\frac1 2 +x)}\,,
\quad
x=\frac{\al} 2+\frac{1-\nu}{2\nu}\,.\label{Calnu}
\end{align}


\section{Level 2}\label{l2}
\subsection{Reflection relations} It has been shown in \cite{NS} that modulo action of the integrals of motion
our fermions solve the reflection relations \cite{FFLZZ}. The goal of the present paper is to incorporate the
descendants created by the integrals of motion into this picture.

We have two reflections:
$$\sigma_1:\ \al\to-\al\,,\qquad \sigma_1:\ \al\to 2-\al\,.$$
Vaguely, the main property of our fermions is that under both of these reflections
act on them as anti-automorphisms with
$$\betab^*_{2j-1}\leftrightarrow\gammab^*_{2j-1}\,.$$
Let us explain this in some more details on a simple example of level 2.

We consider  CFT with the central charge
$$c=1+6Q^2\,,$$
in our notations
$$Q^2=\frac{\nu^2}{\nu-1}\,.$$
To simplify comparison with  \cite{NS} we quote \cite{HGSV}
$$\nu=1+b^2,\quad \al\nu=2 a b\,.$$

In the UV limit the field $\varphi(z,\bar z)$ splits into two chiral components $\phi(z),\phi(\z)$. Consider one of them.
It can be written using the Heisenberg generators and zero mode:
$$\phi(z)=\phi_0-2i\pi _0\log(z)+i\sum_{k\in \mathbb{Z}\backslash 0}^{\infty}\frac {a_k} kz^{-k}\,,$$ 
where
$$[a_k,a_l]=2k\delta_{k,-l}\,,\quad \pi_0=\frac{\partial}{i\partial\phi_0}\,.$$
Define the Virasoro generators
\begin{align}&\mathbf{l}_{k}=\frac 1 4\sum_{j\ne 0,k}a_{j}a_{k-j}+  (i(k+1)Q/2
+\pi _0)a_{k}\,,\quad k\ne 0\,,\label{vir}\\
&\mathbf{l}_{0}=\frac1 2\sum_{j=1}^{\infty}a_{-j}a_j+\pi_0(\pi_0+iQ)\,.\nn
\end{align}
We consider the Heisenberg and Virasoro descendants of the primary field $\Phi_\al(0)$.
The rule of the game is that the Heisenberg descendants are invariant under $\sigma_1$
and the Virasoro descendants are invariant under $\sigma_2$. The goal is to
find a basis, invariant under both. The integrals of motion are invariant under both reflections
by their construction. Up to the level 5 they are
\begin{align}
&\mathbf{i}_1=\mathbf{l}_{-1}\,,\qquad\mathbf{i}_3=2\sum\limits_{k=-1}^{\infty}\mathbf{l}_{-3-k}\mathbf{l}_{k}\,,\nn\\
&\mathbf{i}_5=3\Bigl(\sum\limits_{k=-1}^{\infty}\sum\limits_{l=-1}^{\infty}
\mathbf{l}_{-5-k-l}\mathbf{l}_{l}\mathbf{l}_{k}
+\sum\limits_{k=-\infty}^{-2}\sum\limits_{l=-\infty}^{-2}
\mathbf{l}_{l}\mathbf{l}_{k}\mathbf{l}_{-5-k-l}\Bigr)
+\frac{c+2} 6
\sum\limits_{k=-1}^{\infty}(k+2)(k+3)\mathbf{l}_{-5-k}\mathbf{l}_{k}\,.\nn
\end{align}

Consider the identity
\begin{align}
\Bigl(\mathbf{l}_{-2}-\frac{\al+1}\al\mathbf{i}_1^2\Bigr)\Phi_\al={\textstyle\frac 1 4}Q^2\cdot
\(\al+{\textstyle\frac 1 \nu}\)\(\al-{\textstyle\frac {1-\nu} \nu}\)a_{-1}^2\Phi_\al\,.\label{VH}
\end{align}
This identity is simple, but it explains how we proceed in general case: we take the
Heisenberg part form \cite{NS}, then by definition it is equal to a vector  obtained by action of Virasoro
generators with even indices only plus descendants of local integrals of motion.
The latter terms is what we are interested in this paper. 
 
Multiply \eqref{VH} by $D_1(\al,\nu)D_1(2-\al,\nu)$with
\begin{align}
D_{2n-1}(\al,\nu)=s(\nu)^{2n-1}\sqrt{\frac{i}{\nu}}\frac 1 {(n-1)!}\frac{\Gamma\(\frac \al 2+\frac{2n-1}{2\nu}\)}{\Gamma\(\frac {\al} 2+\frac{(2n-1)(1-\nu)}{2\nu}\)}\,,\nn
\end{align}
obtaining
\begin{align}
&D_1(\al,\nu)D_1(2-\al,\nu)\Bigl(\mathbf{l}_{-2}-\frac{\al+1}\al\mathbf{i}_1^2\Bigr)\Phi_\al\label{id2}\\&={\textstyle\frac 1 4}Q^2\cdot
D_1(\al,\nu)D_1(2-\al,\nu)\(\al+{\textstyle\frac 1 \nu}\)\(\al-{\textstyle\frac {1-\nu} \nu}\)a_{-1}^2\Phi_\al\,.\nn
\end{align}
The right hand side is  invariant under $\sigma_1$. The first term of the left hand
side is invariant under $\sigma_2$, but the second is not. Let us correct this adding $x(\al)\mathbf{i}_1^2\Phi_\al$.

We do not want to spoil the invariance of the right hand side, so, we require
\begin{align}
x(\al)=x(-\al)\,.\label{eqx1}
\end{align}
On the other hand we want to correct the invariance of the left hand side under $\sigma_2$ which
requires
\begin{align}
x(\al)-x(2-\al)=D_1(\al,\nu)D_1(2-\al,\nu)\cdot\frac {2 (1-\al)}{\al(2-\al)}\,.\label{eqx2}
\end{align}

\subsection{Fermionic basis on level 2}
According to our logic the fermionic basis on level 2 (which consists of one vector) is obtained
solving \eqref{eqx1}, \eqref{eqx2}. Solutions are defined up to arbitrary even and periodic with
period $2$ function of $\al$. Let us make the minimality assumption:
\begin{itemize}
\item There are no singularities in the strip $0<\mathrm{Re}(\al)<2$.
\item There is no growth for $\mathrm{Im}(\al)\to \pm\infty$. 
\end{itemize}
For the moment we cannot justify these assumptions, but they supported by extensive numerical 
study as will be explained later. 

With the above assumptions we find
\begin{align}
\betab^*_1\gammab^*_1\Phi_\al=\Bigl(D_1(\al,\nu)D_1(2-\al,\nu)\Bigr(\mathbf{l}_{-2}-\frac{\al+1}{\al}\mathbf{i}_1^2
\Bigl)+\Bigl(A_{1,1}(\al,\nu)+B_{1,1}(\nu)\Bigr)\mathbf{i}_1^2\Bigr)\Phi_\al\,.\label{FIN11}
\end{align}
The function  $A_{1,1}(\al,\nu)$ is defined for $0<\al<1$ by the integral
\begin{align}
&{s(\nu)^{-2}}A_{1,1}(\al,\nu)=\sin\pi\Bigl(\frac{1-\nu}{2\nu}+\frac \al 2  \Bigr)\sin\pi\Bigl(\frac{1-\nu}{2\nu}-\frac \al 2  \Bigr)\nn\\&\times\frac{i}{2\pi^2\nu}\int\limits_{-\infty}^{\infty}\tanh\frac{\pi}2(t+i\al)\left|
\Gamma\Bigl(\frac{2\nu-1}{2\nu}+\frac{it}2\Bigr)
\Gamma\Bigl(\frac{\nu+1}{2\nu}+\frac{it}2\Bigr)
\right|^2\frac{t}{t^2+1}dt\nn\,,
\end{align}
then it is continued analytically. We have intentionally chosen $A_{1,1}(\al,\nu)$ in such a way that
\begin{align}
A_{1,1}({\textstyle\frac{1-\nu}\nu},\nu)=0\,.\label{reqA}
\end{align}
Observe that for $\al =\frac {1-\nu}\nu$ the left hand side of \eqref{id2} is a singular vector.
Then the compatibility with \eqref{sing12} imposes for the constant $B_{1,1}(\nu)$:
\begin{align}
B_{1,1}(\nu)=\frac i {8\nu}
\cot\pi\({\textstyle \frac{1-\nu} {2\nu}}\)C_1(\nu)^2\,.\nn
\end{align}

Now comes the first crucial check of our construction. There is one more singular vector
which contains only $\betab^*_1\gammab_1^*$, this is \eqref{sing13}:
\begin{align}
\Psi_{1,3}&=\[\cot {\textstyle\frac{\pi}2 \(\frac{1-\nu}\nu\)}\taub^*_1\betab^*_1\gammab^*_1+\frac i {\nu}
\taub^*_3
-\frac i {48\nu }\(1+ 3\cot {\textstyle\frac{\pi}2 \(\frac{1-\nu}\nu\)}^2\)(\taub^*_1)^3\]\Phi_{1,3}\,,\label{sing13x}
\end{align}
in which we substitute $\betab^*_1\gammab^*_1\Phi_{1,3}$ given by 
\eqref{FIN11}.
 It is easy to find the
singular vector in terms of Virasoro generators:
\begin{align}
\Psi_{1,3}&=\cot {\textstyle\frac{\pi}2 \(\frac{1-\nu}\nu\)}C_1(\nu)D_1(2{\textstyle\frac{1-\nu}\nu},\nu)
D_1(2-2{\textstyle\frac{1-\nu}\nu},\nu)\label{sing13y}\\&\times
\Bigl(\mathbf{i}_1\mathbf{l}_{-2}- \frac{3 - 2 \nu}{6 (1 - 2 \nu)}\mathbf{i}_3+
 \frac{\nu}{3 (1 -\nu) (1 - 2 \nu)}\mathbf{i}_1^3
\Bigr)\Phi_{1,3}\,,\nn
\end{align}
where the overall
multiplier is chosen  in order to equalise the coefficient in front of $\mathbf{i}_1\mathbf{l}_{-2}$
in \eqref{sing13x} and \eqref{sing13y}.
Then
the coefficient in front of $\mathbf{i}_3$ is automatically consistent while the consistency of
the coefficient in front of $\mathbf{i}_1^3$ requires the identity:
\begin{align}
A_{1,1}\( {\textstyle\frac{2(1 - \nu)}\nu}, \nu\)&=\frac{ (3 - 2 \nu) (2 - 3 \nu)}{
   6 (1 - \nu) (1 - 2 \nu)} D_1\(   {\textstyle\frac{2(1 - \nu)}\nu}, \nu\)
 D_1\(2-   {\textstyle\frac{2(1 - \nu)}\nu}, \nu\)   \nn\\&
    + 
  \frac i{ 48\nu} \(\tan {\textstyle\frac\pi 2 \(\frac {1 - \nu}\nu\)} - 3\cot {\textstyle\frac\pi 2 \(\frac {1 - \nu}\nu\)}  \)
    C_1(\nu)^2\,.\nn
\end{align}

For this identity is quite non-trivial we
write it in the most explicit form starting from $2/3<\nu<1$ (when $\frac{2(1-\nu)}{\nu}<1$) and then continuing
analytically in $\nu$:
\begin{align}
&
4\nu^2\int\limits_{-\infty}^{\infty}\tanh\frac{\pi}2\(t+i{\textstyle\frac{2(1-\nu)}{\nu}}\)\left|
{\textstyle \Gamma\Bigl(\frac{2\nu-1}{2\nu}+\frac{it}2\Bigr)
\Gamma\Bigl(\frac{\nu+1}{2\nu}+\frac{it}2\Bigr)}
\right|^2\frac{t}{t^2+1}dt\label{id13}
\\
&
=\frac {\pi (1 - \nu) }{3\sin \pi\(\frac{1-\nu}{\nu}\)} 
    { \Gamma\({\textstyle\frac 1{2 \nu}}\)^2}
    { \Gamma\(-{\textstyle \frac{1 - \nu}{2 \nu}}\)^2}\nn\\&  
    +
 \frac{2 - 3 \nu}{3 - 2 \nu}\Gamma\(- {\textstyle\frac1{2 \nu}}\)\Gamma\( {\textstyle\frac 3{2 \nu}}\)
    \Gamma\( {\textstyle\frac{1 - \nu}{2 \nu}}\) \Gamma\(-{\textstyle\frac 3 2\(\frac{1 - \nu}{2 \nu}\)}\)
    \cdot
\left\{  \begin{matrix}
  {3 - 2 \nu}\,,&\quad \nu>2/3\\ 
 {3 - 5 \nu} \,,&\quad \nu<2/3
    \end{matrix}\right.
    \,.\nn
\end{align}
We do not have a good analytical proof of this identity, even computing by residues is hard
because double poles are present, but numerical check supports it perfectly.

\section{ Level 4}\label{l4}

Following \cite{NS} consider two vectors in Heisenberg representation
\begin{align}
&H_{1,3}=\frac 1{432 (1 - \nu)^3} (\al \nu+1) ( \al \nu-3(1-\nu))\Bigl(U+\al \nu(2-\nu)  W \Bigr)\,,\nn\\
   &H_{3,1}=\frac 1{432 (1 - \nu)^3} ( \al \nu+3) ( \al \nu+\nu-1) \Bigl(U-\al \nu(2-\nu)  W \Bigr)\,,\nn  
\end{align}
where
\begin{align}
&U=3 \Bigl( 6 \nu - 3 \nu^2 + 2 \al^2 \nu^2 - 2 \al^2 \nu^3 + \al^2 \nu^4-3)a_{-1}^4+12 (1 - \nu) (1 - \nu -\nu^2)a_{-2}^2 \Bigr)\,,\nn\\
&W=(3 - 3 \nu + \al^2 \nu^2) a_{-1}^4 -12 (1 - \nu) a_{-2}^2\,.\nn
\end{align}
Rewrite them in Virasoro representation
\begin{align}
&H_{1,3}=V_{1,3}\,,\quad  
   H_{3,1}=V_{3,1}\,,\nn  
\end{align}
\begin{align}
V_{1,3}&=\Bigr(\mathbf{l}_{-2}^2+\Bigl(\frac{2c-32}9+\frac 2 3 d(\al)\Bigr)\mathbf{l}_{-4}
+X_{1,3}^{1,3}(\al,\nu)\mathbf{i}_1\mathbf{i}_3\label{final13}\\&+X_{1,3}^{1,1|1,1}(\al,\nu)\mathbf{i}_1^2\Bigr(\mathbf{l}_{-2}-\frac{\al+1}{\al}\mathbf{i}_1^2\Bigl)\mathbf{i}_1^2
+X_{1,3}^{1,1,1,1}(\al,\nu)\mathbf{i}_1^4
\Bigr)\,,\nn\\V_{3,1}&=\Bigr(\mathbf{l}_{-2}^2+\Bigl(\frac{2c-32}9-\frac 2 3 d(\al)\Bigr)\mathbf{l}_{-4}
+X_{3,1}^{1,3}(\al,\nu)\mathbf{i}_1\mathbf{i}_3\nn\\&+X_{3,1}^{1,1|1,1}(\al,\nu)\mathbf{i}_1^2\Bigr(\mathbf{l}_{-2}-\frac{\al+1}{\al}\mathbf{i}_1^2\Bigl)
+X_{3,1}^{1,1,1,1}(\al,\nu)\mathbf{i}_1^4
\Bigr)\,,\nn
\end{align}
where
$$d(\al)=\frac{(1- \al) (2 - \nu) \nu}{1 - \nu}\,,$$
The coefficients $X(\al,\nu)$
are rational functions of their arguments, they  are given in the Appendix. The important information about them is that
in the domain $0<\mathrm{Re}(\al)<2$ all of them have simple pole at $\al=2\frac{1-\nu}\nu$, and
$X_{1,3}^{1,1|1,1}(\al,\nu)$ has additional simple pole at $\al=\frac{1-\nu}\nu$.

We look for the fermionic basis on level four in the form
\begin{align}
 \betab^*_1\gammab^*_3\Phi_\al&=\Bigl[\half
 D_1(\al,\nu)D_3(2-\al,\nu)V_{1,3}\label{FIN13}\\&+\Bigl(A_{1,3}(\al,\nu)+B_{1,3}(\nu))\Bigr)\mathbf{i}_1\mathbf{i}_3\,,\nn\\
&+\Bigl(A_{1,1|1,1}(\al,\nu)+B_{1,1|1,1}(\nu)\Bigr)\mathbf{i}_1^2\betab^*_1\gammab^*_1
\nn\\&+\Bigl(A_{1,1,1,1}(\al,\nu)+B_{1,1,1,1}(\nu))\Bigr)\mathbf{i}_1^4\Bigr]\Phi_\al\,,\nn
\end{align}
\begin{align}
 \betab^*_3\gammab^*_1\Phi_\al&=\Bigl[\half
 D_3(\al,\nu)D_1(2-\al,\nu)V_{3,1}\label{FIN31}\\&+\Bigl(A_{1,3}(-\al,\nu)+B_{1,3}(\nu))\Bigr)\mathbf{i}_1\mathbf{i}_3\,,\nn\\
&+\Bigl(A_{1,1|1,1}(-\al,\nu)+B_{1,1|1,1}(\nu)\Bigr)\mathbf{i}_1^2\betab^*_1\gammab^*_1
\nn\\&+\Bigl(A_{1,1,1,1}(-\al,\nu)+B_{1,1,1,1}(\nu))\Bigr)\mathbf{i}_1^4\Bigr]\Phi_\al\,.\nn
\end{align}
The functions $A(\al,\nu)$ should satisfy difference equations similar to \eqref{eqx2}. To all of them
we impose the normalisation
\begin{align}
A_{\#}({\textstyle 3\frac{1-\nu}\nu},\nu)=0\,,\label{norm13}
\end{align}
in order to be able to normalise by the singular vector of $\Phi_{1,4}$.
We require that \eqref{FIN13}, \eqref{FIN31} are regular for $0<\al<2$.
We suppose that the procedure is clear now, so, without going into much details we give the results.

The coefficients $A$ are obtained by analytical continuation from the domain $1/2\le\nu<2/3$, $0<|\al|<1$ of
the following integrals.
\begin{align}
{s(\nu)^{-4}}A_{1,3}(\al,\nu)&=-\frac{ 1} {8 \nu \pi^2}\int\limits_{-\infty}^{\infty} 
\frac{\sin\frac\pi 2 \(\al-3\frac{1-\nu}\nu\)\sinh\frac\pi 2 \(t-\frac i \nu\)}{\cosh\frac\pi 2(t+i\al)}
\label{A13}\\&\times
     {\textstyle \Gamma\(1 - \frac 1{2 \nu} - \frac{i t}2\) \Gamma\(2 - \frac3 {2 \nu} + \frac{i t}2\)
    \Gamma\(
 \frac{ 3 + \nu}{2 \nu} -\frac{ i t}2 \) \Gamma\(\frac{1 +\nu}{2\nu} + \frac{i  t} 2 \)}\nn\\&\times
 \bigl(X_{3,1}^{1,3}(1 - it, \nu) -X_{1,3}^{1,3}(1 + it, \nu) \bigr)dt\nn\\
&
      - \frac{ 3 i \sqrt{\pi}
  \Gamma\(-\frac 3 2 + \frac 2 \nu\)}{8 \nu \Gamma\(-2 + \frac 2 \nu\)}
  \bigl(X_{3,1}^{1,3}(2-3{\textstyle\frac {1-\nu}\nu}, \nu) -X_{1,3}^{1,3}(3{\textstyle\frac {1-\nu}\nu}, \nu) \bigr) \,.\nn
\end{align}
Here the last term is added because the integral does not vanish at $\al=3\frac{1-\nu}\nu$. Indeed, in the range of $\nu$ which
we consider for this value of $\al$ the pole of the integrand at $i(1-\al)$ crosses the real line and punches the integration contour 
together with the pole at $t=i(4-3/\nu)$ of the second gamma-function. Because of the multiplier
$\sin\frac\pi 2 \(\al-3\frac{1-\nu}\nu\)$ the integral at this value of $\al$ is finite, it is cancelled by the last term in \eqref{A13}.

\begin{align}
{s(\nu)^{-2}}A_{1,1|1,1}(\al,\nu)&=\frac 1 {8 \pi^2}\int\limits_{-\infty}^{\infty} \frac{\sin\frac\pi 2 \(\al-3\frac{1-\nu}\nu\)
\cosh\frac\pi 2 \(t+\frac i \nu\)}{\cosh\frac\pi 2(t+i\al)}\label{A11|11}\\&\times
  {\textstyle \Gamma\(
  2 - \frac3 {2 \nu} + i\frac  t 2\) \Gamma\(\frac 1{2 \nu} -  i\frac  t 2\) \Gamma\(
  \frac{3 + \nu }{2 \nu}- i\frac  t 2\) \Gamma\(-\frac{1 - \nu }
  {2 \nu}+ i\frac  t 2\)}\nn\\&\times\bigl(X_{3,1}^{1,1|1,1}(1 - i t, \nu) -X_{1,3}^{1,1|1,1}(1+i t, \nu)\bigr)
  dt\nn\\&-\frac{\sin\frac\pi 2\(\al-3\frac{1-\nu}\nu\)}{\sin\(\frac\pi\nu\)\sin\frac\pi 2\(\al-\frac{1-\nu}\nu\)}
  \frac{\sqrt{\pi}\Gamma\(\frac 3 2  + \frac1 \nu\)}{2\Gamma\(\frac 1 \nu\)}\res_{\al=\frac{1-\nu}\nu}
  X_{1,3}^{1,1|1,1}(\al, \nu)
 \,.\nn
\end{align}
Here the last term is added in order to cancel the pole at $\al=\frac{1-\nu}\nu$ of \\ \mbox{
$\frac 1 2 D_1(\al,\nu)D_3(2-\al,\nu)X_{1,3}^{1,1|1,1}(\al,\nu)\,,$}
in \eqref{FIN13}.

\begin{align}
{s(\nu)^{-4}}A_{1,1,1,1}(\al,\nu)&=
\frac 1{4\pi^2}\int\limits _{-\infty}^{\infty}\frac{\sin\frac\pi 2 \(\al-3\frac{1-\nu}\nu\)
\cosh\frac\pi 2 \(t+\frac i \nu\)}{\cosh\frac\pi 2(t+i\al)}\label{A1111}\\&\times
  {\textstyle  \Gamma\(
  2 - \frac3 {2 \nu} + i\frac  t 2\) \Gamma\(\frac 1{2 \nu} -  i\frac  t 2\) \Gamma\(
  \frac{3 + \nu }{2 \nu}- i\frac  t 2\) \Gamma\(-\frac{1 - \nu }
  {2 \nu}+ i\frac  t 2\)}\nn\\&\times\Bigl(
  X_{1,3}^{1,1|1,1}(1+it,\nu)A_{1,3}(1+it,\nu)
  - X_{3,1}^{1,1|1,1}(1-it,\nu)A_{1,3}(1-it,\nu)
  \Bigr)dt\nn\\&-\frac 1 {8\nu\pi^2}\int\limits_{-\infty}^{\infty} \frac{\sin\frac\pi 2 \(\al-3\frac{1-\nu}\nu\)\sinh\frac\pi 2 \(t-\frac i \nu\)}{\cosh\frac\pi 2(t+i\al)}\nn\\&\times{\textstyle
     \Gamma\(1 - \frac 1{2 \nu} - \frac{i t}2\) \Gamma\(2 - \frac3 {2 \nu} + \frac{i t}2\)
    \Gamma\(
 \frac{ 3 + \nu}{2 \nu} -\frac{ i t}2 \) \Gamma\(\frac{1 +\nu}{2\nu} + \frac{i  t} 2 \)}\nn\\&\times
 \bigl(X_{3,1}^{1,1,1,1}(1 - it, \nu) 
 - X_{1,3}^{1,1,1,1}(1+it, \nu)\bigr)dt\nn\\&
  - \frac{ 3 i \sqrt{\pi}
  \Gamma\(-\frac 3 2 + \frac 2 \nu\)}{8 \nu \Gamma\(-2 + \frac 2 \nu\)}  \bigl(X_{3,1}^{1,1,1,1}(2-3{\textstyle\frac {1-\nu}\nu}, \nu)-
 X_{1,3}^{1,1,1,1}(3{\textstyle\frac {1-\nu}\nu}, \nu)\bigr)\,,\nn
\end{align}
where the last term is added in order to cancel the none-zero value of the second integral at
$\al=3(1-\nu)/\nu$ which occurs for the same reason as in \eqref{A13}.

Due to \eqref{norm13} the constants $B$ are read from the singular vector \eqref{sing14}:
\begin{align}
&B_{1,3}(\nu)=\frac i {8\nu}\(\cot {\textstyle\frac{3\pi}2 \(\frac{1-\nu}\nu\)}
+2\cot {\textstyle\frac{\pi}2 \(\frac{1-\nu}\nu\)}\)C_1(\nu)C_3(\nu)\,,\label{Bs}\\
&B_{1,1|1,1}(\nu)=\frac 1 8\cot {\textstyle\frac{\pi}2 \(\frac{1-\nu}\nu\)}^2
C_1(\nu)^2\,,\nn\\
&B_{1,1,1,1}(\nu)=-
\frac i {192\nu}\(\cot {\textstyle\frac{\pi}2 \(\frac{1-\nu}\nu\)}+\cot {\textstyle\frac{\pi}2 \(\frac{1-\nu}\nu\)}^3\)C_1(\nu)^4\,.\nn
\end{align}

These formulae should be checked against the singular vectors \eqref{sing14} (which is satisfied by
definition), \eqref{sing22}, \eqref{sing15} and the descendants \eqref{null12}, \eqref{null11}, \eqref{null115}, \eqref{null13}. This leads to a number of identities for one-fold integrals and two-fold integrals which we 
have verified numerically. 

To give an example let us consider the most non-trivial case 
provided by \eqref{null13}:
\begin{align}
&\Bigr[\cot {\textstyle\frac{\pi}2 \(\frac{1-\nu}\nu\)}\taub^*_1\betab^*_3\gammab^*_1
+\frac i {\nu}\taub^*_5-
\frac i {16\nu}(1+\cot {\textstyle\frac{\pi}2 \(\frac{1-\nu}\nu\)}^2)
\taub^*_3(\taub^*_1)^2\label{xxx1}
\\&+\frac i {1920\nu}(1-5\cot {\textstyle\frac{\pi}2 \(\frac{1-\nu}\nu\)}^2)(\taub^*_1)^5
\Bigl]\Phi_{1,3}\nn\,.
\end{align}
One easily finds the general null-vector of $\Phi_{1,3}$ on level 5. It consists of a non-trivial
one together with the simple descendent created by $\mathbf{i}_1^2$ from the
singular vector with arbitrary coefficient:
\begin{align}
&\Bigl[\mathbf{i}_1\mathbf{l}_{-2}^2-\frac{2 (1 +3 \nu - \nu^2)}{3 (1 - \nu)}\mathbf{i}_1\mathbf{l}_{-4}-\frac{5 - 2 \nu}{15 (1 - 2 \nu)}\mathbf{i}_5+C\cdot\mathbf{i}_1^3\Bigl(\mathbf{l}_{-2}-\frac{2 - \nu}{2 (1 - \nu)}\mathbf{i}_1^2\Bigr)
\label{xxx2}\\&+\frac{2 - (1-\nu)(3-2\nu)C}{6 (1 - \nu) (1 - 2 \nu)}\mathbf{i}_3\mathbf{i}_1^2
-\frac{5 - 3 \nu-5 (1 - \nu) (3 -2 \nu) (2 - 3 \nu)C}{30 (1 - \nu)^2 (1 - 2 \nu)}\mathbf{i}_1^5\Bigr]\Phi_{1,3}\,.\nn
\end{align}
Now we multiply \eqref{xxx2} by $\frac 1 2\cot {\textstyle\frac{\pi}2 \(\frac{1-\nu}\nu\)}C_1(\nu)
D_3(\frac{2(1-\nu)}{\nu},\nu)D_1(2-\frac{2(1-\nu)}{\nu},\nu)$ and compare. The coefficients in front
of $\mathbf{i}_1\mathbf{l}_{-4}$ and  $\mathbf{i}_5$ are all right due to
\begin{align}
&\frac{2c-32} 9-\frac{2}{3}d\bigl({\textstyle\frac{2(1-\nu)}\nu}\bigr)=-\frac{2 (1 +3 \nu - \nu^2)}{3 (1 - \nu)}\,,\nn\\
&\frac i \nu \tan {\textstyle\frac{\pi}2 \(\frac{1-\nu}\nu\)} 
\frac{2C_3(\nu)}{ D_3(\frac{2(1-\nu)}{\nu},\nu)D_1(2-\frac{2(1-\nu)}{\nu},\nu)C_1(\nu)}
=-\frac{5 - 2 \nu}{15 (1 - 2 \nu)}\,.\nn
\end{align}
Let us consider the coefficient of $\mathbf{i}_1^3\mathbf{l}_{-2}$. 
This is not quite trivial because both $X_{3,1}^{1,1|1,1}(\al,\nu)$ and $A_{1,1|1,1}$ develop simple
poles at $\al=2\frac{1-\nu}\nu$, they cancel only in the final expression \eqref{FIN31}. Let us explain
how it happens. In our domain $1/2<\nu<2/3$ the point $-2\frac{1-\nu}\nu$ lies below $-1$, hence the
integral needs to be continued analytically.
Actually, we can use the functional relation 
\begin{align}
A_{1,1|1,1}(-\al,\nu)+\frac12\frac{D_3(\al,\nu)}{D_1(\al,\nu)}
X_{3,1}^{1,1|1,1}(\al,\nu)=A_{1,1|1,1}(2-\al,\nu)+\frac12\frac{D_3(\al,\nu)}{D_1(\al,\nu)}
X_{1,3}^{1,1|1,1}(2-\al,\nu)\label{123a}\,,
\end{align}
in order to compute the constant $C$ because
both terms in the right hand side of \eqref{123a} are regular at $\al=2\frac{1-\nu}\nu$,
\begin{align}
&C=X_{1,3}^{1,1|1,1}\(2-{\textstyle\frac{2(1-\nu)}\nu}\)+
2\frac{D_1\(\frac{2(1-\nu)}\nu,\nu\)}{D_3\(\frac{2(1-\nu)}\nu,\nu\)}
\Bigl(A_{1,1|1,1}\(2-{\textstyle\frac{2(1-\nu)}\nu},\nu\)+B_{1,1|1,1}(\nu)\Bigr)\,.\label{Ca}
\end{align}

Performing a similar computation for the term with $\mathbf{i}_1^2\mathbf{i}_3$ we find
\begin{align}
&X_{1,3}^{1,3}\(2-{\textstyle\frac{2(1-\nu)}\nu},\nu\)+\frac 2 {D_3\(\frac{2(1-\nu)}\nu,\nu\)D_1\(2-\frac{2(1-\nu)}\nu,\nu\)}\Bigl[ A_{1,3}\(2-{\textstyle\frac{2(1-\nu)}\nu},\nu\)+B_{1,3}(\nu)
\nn\\&
{+
\frac i{8\nu\sin \(\frac{\pi}\nu\)}C_1(\nu)C_3(\nu)}
\Bigr]
=-\frac{   (3 - 2 \nu)}{
	6  (1 - 2 \nu)}\cdot C+\frac2{
	6 (1 - \nu) (1 - 2 \nu)}\,,\nn
\end{align}
where $B_{1,1|1,1}$ and $B_{1,3}$ are given by (\ref{Bs}).
Having in mind the formula for $C$ \eqref{Ca} we obtain quite non-trivial identity including the
integrals $A_{1,3}\(2-{\textstyle\frac{2(1-\nu)}\nu},\nu\)$ and 
\newline
$ A_{1,1|1,1}\(2-{\textstyle\frac{2(1-\nu)}\nu},\nu\)$ which we checked numerically.
The coefficient of $\mathbf{i}_1^5$ yields still more complicated identity which includes two-fold and one-fold
integrals. This identity also can be verified numerically.

\section{Expectation values on a cylinder}\label{expectation}

We begin study of the Euclidean sine-Gordon model
on a cylinder of radius $R$ ($(z\in\mathbb{C}/2\pi i R\mathbb{Z}$).
We require that the twists at $\mathrm{Re}(z)\to \infty$ and $\mathrm{Re}(z)\to -\infty$
are respectively $p$ and $p'$. A local operator created by 
our fermions
fermions $\betab^*_{2j-1}$, $\gammab^*_{2k-1}$, $\bar{\betab}^*_{2j-1}$, $\bar{\gammab}^*_{2k-1}$
from the primary field $\Phi_\al$  is inserted at $\z=0$. Before going further let us fix notations.

In what follows we shall use rapidity-like variables
which are related to the original
spectral parameters  as
$$\theta=\frac 1 \nu\log\z\,.$$
The radius of the cylinder (inverse temperature) will be parametrised by dimensionless 
parameter $\theta_0$:
\begin{align}
e^{-\theta_0}=\mub^{\frac 1 \nu} R\,.\label{defth0}
\end{align}
The high temperature limit corresponds to $\theta_0\to\infty$.

General formula \cite{HGSII,HGSV} expresses the expectation values in terms of two functions
$\Omega(\theta,\sigma|\al,\theta _0)$ and $\varrho(\theta|\theta_0)$   (mostly we omit the dependence
upon $p,p'$, but when needed we make it explicit writing $\Omega(\theta,\sigma|\al,\theta _0,p,p')$ {\it etc}).
For $p=p'$ the function $\Omega(\theta,\sigma|\al,\theta _0)$ is related to the one used in \cite{HGSV} via
$$\Omega(\theta,\sigma|\al,\theta_0)=\omega_R^\mathrm{sG} \bigr(e^{\nu\theta},e^{\nu\sigma}|\al\bigl)\,.$$
{The main goal of this  section is to show how our previous results allow to compute the high temperature
limit for the asymptotical coefficients of $\Omega(\theta,\sigma|\al,\theta _0)$.}

Before writing down the formulae {for the expectation values}  let us comment on their origin.
Our treatment of the sG model is based on the scaling limit of
an inhomogeneous six-vertex model, as explained in \cite{HGSV}. If $p'=p+\al\nu/2$ the 
expectation value in question allows simple interpretation as scaling limit of the 
one computed in \cite{HGSIV}, $\kappa=2p/\nu$ and $\kappa'=\kappa+\al$ being the twists of the
Matsubara transfer-matrices to the left and to the right of the inserted local field.
More sophisticated construction including the lattice screening operators \cite{HGSIV,OP,HGSV}
allows to consider $p'=p+\al\nu/2+(1-\nu)m$ for $m=0,1,2,\cdots$. In all these cases the
expectation values are expressed in terms of the function $\Omega(\theta,\theta'|\al,\theta_0)$. This
function is defined through certain integral equation, and 
the only difference for different boundary conditions is that we have to insert the ratio of
eigenvalues of Matsubara transfer-matrices for $p'$ and $p$  into the kernel {of an integral equation
(see formula
\eqref{measure} below)}. This makes us to conjecture that arbitrary values of $p$, $p'$
can be considered in the same fashion. 

Let us make one important remark. The twists $p$, $p'$ enter quite
differently 
the equation for the function $\Omega(\theta,\theta'|\al,\theta_0)$ which we shall  formulate.
However, the final results must be symmetric with respect to interchange $p\leftrightarrow p'$.
We shall check that this is actually the case. This check provides  a crucial support to our procedure.

General formula is only slightly different from the one given in \cite{HGSV}, so we shall be brief.
As usual we introduce the multi-index notations for products of  fermions. Then
\begin{align}
&\frac {\langle
\betab^* _{I^+}\bar{\betab} ^*_{\bar{I}^+}
\bar{\gammab }^*_{\bar{I}^-}\gammab^*_{I^-}
\Phi _{\al}(0)
 \rangle}
{\langle\Phi_{\al }(0) \rangle}\label{themain}
= \mathcal{D}\(I^+\cup (-\bar{I}^+)\ |\ I^-\cup
(-\bar{I}^-)\)\,,
\end{align}
implying $\#(I^+)+\#(\bar{I}^+)=\#(I^-)+\#(\bar{I}^-)$, $\mathcal{D}$ stands for the determinant
defined for two sets of odd integers $A=\{a_1,\cdots, a_n\}$, $B=\{b_1,\cdots, b_n\}$
\begin{align*}
&\mathcal{D}(A|B)=\prod\limits _{j=1}^n
\mathrm{sgn}(a_j)\mathrm{sgn}(b_j)\det\( D_{a_n,b_k}(\al)\)|_{j,k=1,\cdots ,n},\\
&D_{a,b}=\Omega_{a,b}(\al,\theta_0)
+\delta_{a,-b}\mathrm{sgn}(a)\Omega_{a,b}^{(0)}(\al,\theta_0)\,,
\end{align*}
Here $\Omega_{2j-1,2k-1}(\al,\theta_0)$ are coefficient of the asymptotics 
a function which will be discussed in the next section
$$
\Omega(\theta,\sigma|\al,\theta_0)\ \ \simeq\hskip -.8cm\raisebox{-.5cm}{\scalebox{1.2}{$
{{\sigma\to
\epsilon\infty}}\atop{{\theta\to
\epsilon'\infty}}$}}\ \ \sum\limits_{j,k=1}^{\infty}e^{-\epsilon(2j-1)\theta}e^{-
\epsilon'(2k-1)\sigma}\Omega_{\epsilon(2j-1),\epsilon'(2k-1)}(\al,\theta_0)\,.
$$
We shall need also the function $\rho(\theta|\theta_0)$
which is the ratio of the ground state Matsubara transfer-matrices:
$$\Rho(\theta|\theta_0,p,p')=\frac{T(\theta|\theta_0,p')}{T(\theta|\theta_0,p,)}\,.$$

{The coefficients $\Omega^{(0)}_{2j-1,2k-1}(\al,\theta_0)$ do not vanish only if
$\sgn(2j-1)\ne \sgn(2k-1)$. They are obtained 
from the expansion for $\theta\to\pm\infty$,
$\sigma\to \mp\infty$ of}
\begin{align}
\Omega_0(\theta,\sigma|\al)&=\frac 1 2\Bigl(\Rho(\theta|\theta_0)^{\frac 1 2}\Rho(\sigma|\theta_0)^{\frac 1 2}\tau_\pm(e^{\sigma-\theta}|\al)
-\Rho(\theta|\theta_0)^{-\frac 1 2}\Rho(\sigma|\theta_0)^{\frac 1 2}\tau_\pm(-e^{\sigma-\theta}|\al)\nn\\&
-\Rho(\theta|\theta_0)^{\frac 1 2}\Rho(\sigma|\theta_0)^{-\frac 1 2}\tau_\pm(-e^{\sigma-\theta}|\al)
+\Rho(\theta|\theta_0)^{-\frac 1 2}\Rho(\sigma|\theta_0)^{-\frac 1 2}\tau_\pm(e^{\sigma-\theta}|\al)\Bigr)\,,\nn
\end{align}
which follows from the asymptotical expansions for $\varrho(\theta|\theta_0)$ given in terms
of  the differences of the of the Mastubara eigenvalues of the
local integrals of motion:
\begin{align}
&\log\varrho(\theta|\theta_0,p,p')\ \ \simeq\hskip -.8cm\raisebox{-.3cm}{\scalebox{.75}{${\theta\to\infty}$}}\ \ \sum_{j=1}^{\infty}C_{2j-1}(\nu)\bigl(I_{2j-1}(\theta_0,p')-I_{2j-1}(\theta_0,p)   \bigr)e^{-(2j-1)(\theta-\theta_0)}\,,\label{asrho}\\
&\log\varrho(\theta|\theta_0,p,p')\ \ \simeq\hskip -.8cm\raisebox{-.3cm}{\scalebox{.75}{${\theta\to-\infty}$}}\ \ \sum_{j=1}^{\infty}C_{2j-1}(\nu)\bigl(\overline{I}_{2j-1}(\theta_0,p')-\overline{I}_{2j-1}(\theta_0,p)   \bigr)e^{(2j-1)
{(\theta+\theta_0)}}\,.\nn
\end{align}
In the high temperature limit
$$I_{2j-1}(\theta_0,p)\ \ \to\hskip -.8cm\raisebox{-.3cm}{\scalebox{.75}{${\theta_0\to\infty}$}}I_{2j-1}(p)\,,\quad
\overline{I}_{2j-1}(\theta_0,p)\ \ \to\hskip -.8cm\raisebox{-.3cm}{\scalebox{.75}{${\theta_0\to\infty}$}}I_{2j-1}(p)\,,
$$
where $I_{2j-1}(p)$ are the CFT eigenvalues \cite{BLZ}. We shall need the first three of them
\begin{align}
&I_1(p)= \frac{p^2}{1 - \nu} - \frac 1 {24}\,,\label{eigenI}\\
&
I_3(p)= I_1(p)^2 - \frac1 6 I_1(p) + \frac{c}{1440}\nn\\
&I_5(p)= 
  I_3(p)I_1(p)- 
   \frac1 3 I_3(p) + \frac{c + 5}{360}I_1(p)- 
  \frac{ c(5 c + 28)}{181440}\,.\nn
\end{align}

For $j,k>0$ in the high temperature limit 
$$e^{-2(k+j-1)\theta_0}\Omega_{2j-1,2k-1}(\al,\theta_0)\ \ \to\hskip -.8cm\raisebox{-.3cm}{\scalebox{.75}{${\theta_0\to\infty}$}} \Omega_{2j-1,2k-1}(\al)\,.$$
We read 
$\Omega_{2j-1,2k-1}(\al,\theta_0)$ 
from 
the particular case of \eqref{themain}
\begin{align}
&\frac {\langle p|\betab^*_{2j-1}\gammab^*_{2k-1}\Phi_\al(0)|p'\rangle} {\langle p|\Phi_\al(0)|p'\rangle}=
\Omega_{2j-1,2k-1}(\al|\theta_0)\,.\nn
\end{align}
and the ``theoretical prediction" for high temperature limits
$\Omega_{1,1}(\al)$, $\Omega_{1,3}(\al)$,
$\Omega_{3,1}(\al)$
from
\eqref{FIN11}, \eqref{FIN13}, \eqref{FIN31}. First, we replace 
$\mathbf{i}_{2j-1}$ by $I_{2j-1}(p') -I_{2j-1}(p)$. Second, we replace the operators $\mathbf{l}_{-2}\Phi_\al$, $\mathbf{l}_{-2}^2\Phi_\al$, $\mathbf{l}_{-4}\Phi_\al$
by their normalised expectation values which are correspondingly
\begin{align}
&\frac 1 2  \bigl(I_1(p)+I_1(p')\bigr) - \frac 1 {12} \Delta_\al\,,\nn\\
&\frac 1 {1440}\Bigl(c + 360 \bigl(I_1(p)+I_1(p')\bigr)^2 + 
   28 \Delta_\al - 120 \bigl(I_1(p)+I_1(p')\bigr) (\Delta_\al+1) + 
   10 \Delta_\al^2\Bigr)\,,\nn\\
   &\frac 1 {240}\Delta_\al\,.\nn
\end{align}
In the next section we shall compare this prediction with TBA-like method for computing $\Omega(\theta,\sigma|\al,\theta_0)$.

For the second chirality the formulae are quite similar, except that we have to
take the asymptotics at $\sigma,\theta\to-\infty$. It is getting more interesting when we mix
two chiralities. Consider, for example, $\betab_1^*\bar{\gammab}^*_1\Phi_\al(0)$. We have
from \eqref{themain}
\begin{align}
&\frac {\langle p|\betab^*_{1}\bar{\gammab}^*_{1}\Phi_\al(0)|p'\rangle} {\langle p|\Phi_\al(0)|p'\rangle}=
\Omega_{1,-1}(\al,\theta_0)+\Omega^{(0)}_{1,-1}(\al,\theta_0)\,.\label{sss}
\end{align}
then using \eqref{shift1} we obtain
\begin{align}
\Omega_{1,-1}(\al,\theta_0)+\Omega^{(0)}_{1,-1}(\al,\theta_0)=\frac{\langle p|\betab_1^*\bar{\gammab}^*_1\Phi_\al(0)|p'\rangle}
{\langle p|\Phi_\al(0)|p'\rangle}=
\frac  {2t_1(\al)}{C(\al,\nu)}
\frac{\langle p|\Phi_{\al+2\frac{1-\nu}\nu}(0)|p'\rangle}{\langle p|\Phi_\al(0)|p'\rangle}\,.\nn
\end{align}
$\Omega^{(0)}_{1,-1}(\al|\theta_0)$ can be evaluated explicitly. The ratio in the right hand side
approaches in the high temperature
limit the ratio of three-point functions for which we use the formulae from \cite{ZZ}. Having all that in mind
we derive ``theoretical" prediction for the high temperature behaviour of $\Omega_{1,-1}(\al|\theta_0)$:
\begin{align}
 e^{-2\theta_0} &\Omega_{1,-1}(\al|\theta_0)\ \ \to\hskip -.8cm\raisebox{-.3cm}{\scalebox{.75}{${\theta_0\to\infty}$}} \label{3point}\\&
\ \ \ \frac i {8\nu} \cot \({\textstyle \frac {\pi}{2 \nu}}\al\) C_1(\nu)^2
           (I_1(p')-I_1(p))^2\qquad\qquad\qquad\qquad\qquad\qquad\ \  (I)\nn\\&-\frac {i}\nu \cot {\textstyle \frac {\pi}{2 \nu}} (\al\nu + 1)e^{-2\theta_0}   \qquad\qquad\qquad\qquad\qquad\qquad\qquad\qquad\qquad\ \  (II)\nn\\&+
           \frac {i}\nu \cot {\textstyle \frac {\pi}{2 \nu}} (\al\nu + 1)
\frac{C(p,p',\al,\nu)}{C(\al,\nu)}
      e^{2 ( 
           \Delta_{\al}-\Delta_{\al + 2 \frac{1 - \nu}{\nu}} -1 )\theta_0 } \,,\qquad\qquad\qquad\ (III)\nn
 \end{align}
the constant $C(\al,\nu)$ is given by \eqref{Calnu},
and $C(p,p',\al,\nu)$ is the ratio of CFT three-point functions:
\begin{align}
C(p,p',\al,\nu)=\Gamma(\nu)^2\gamma(\nu(1-\al))\gamma(1+\nu(2-\al))
\prod_{\epsilon,\epsilon '=\pm}\gamma(1+\nu\al/2-\nu+\epsilon p+\epsilon 'p') \,,
\end{align}
where $\gamma(x)=\Gamma(x)/\Gamma(1-x)$.
This is the simplest formula including shifts,
but on the other hand this is the most important one: it is sufficient for the study of
ultra-violet behaviour of sG correlation functions in the framework of perturbed CFT. 
We counted three different parts of the  right hand ride of \eqref{3point} 
by Roman numerals for future use. 

\section{Numerical work}\label{numerics}

We proceed to the numerical check of our results.  This  requires knowledge of the 
Matsubara
ground state
as function of twist. This part of the problem is well-known, but we give corresponding
equations in order to fix the notations. We also provide some comments on our tricks (probably not the most
intelligent ones) for doing numerics on Mathematica.

We introduce in usual way the auxiliary function $\mathfrak{a}(\theta|\theta_0,p)$, and write the non-linear 
equation \cite{BKP,DDV}:
\begin{align}
&\log\mathfrak{a}(\theta+i\eta|\theta_0,p)=2\pi i  MR\sinh(\theta+i\eta)-\frac{2\pi i }  {1-\nu} p\label{DDV}\\&-\int_{-\infty}^{\infty}
\Bigl(G(\theta-\theta')\log\(1+\mathfrak{a}(\theta'+i\eta|\theta_0,p)\)-G(\theta-\theta'+2i\eta)
\log\(1+\overline{\mathfrak{a}(\theta'+i\eta|\theta_0,p))}\)d\theta'\,,\nn
\end{align}
where $M$ is the soliton mass which we relate to $\theta_0$ by well-known
 formula \cite{Alyoshascale}:
$$MR=\Gamma(\nu)^{\frac 1 \nu} \frac {2 \Gamma\(\frac{1 - \nu}{2 \nu}\)}{\sqrt{\pi}
      \Gamma\(\frac{1}{2 \nu}\)}e^{-\theta_0}\,.$$
The kernel is 
\begin{align}
G(\theta)=\int_{-\infty}^{\infty}\frac{\sinh\frac {\pi k} 2\(\frac{2\nu-1}{\nu}\)}
{4\pi \sinh\frac {\pi k} 2\(\frac{1-\nu}{\nu}\)\cosh\(\frac {\pi k} 2\)}e^{ik\theta }dk\,.
\end{align}Mathematica is not very strong in Fourier transform, we do the integral and compare it with
first 30 terms of the asymptotics, when the difference is $10^{-20}$ (this normally happens for $-2<\theta<2$)
we stop computing the integral and switch to the asymptotics. In this way we avoid rapidly oscillating integrals. 
Finally, the number $\eta$ is rather arbitrary, from $0$ to $\pi\frac{1-\nu}{2\nu}$. It should not be too close to 
the minimal and maximal values. Our preferable choice is $\eta =\pi\frac{2(1-\nu)}{5\nu}$, but for reasons which
will be clear later we need one more value of $\eta$. In this case we take $\eta =\pi\frac{1-\nu}{5\nu}$. This is
already close to $0$, that is why in the first case we replace the integral by sum with the step $1/10$, but and in the second with the step $1/20$.

In order to be able to compare with CFT we are interested in the high temperature behaviour, in other words in
small $R$. The values of $\theta_0$ of the order $10$ work perfectly. Here is the graphic of the real  
and imaginary parts
of $\log\(1+\mathfrak{a}\(\theta+\pi i\eta \)\)$ ($\eta= \frac {2(1-\nu)} {5\nu}$) for $p=0$, $\nu=5/8$, and
$\theta_0=10,12$ which shows well
pronounced  splitting into two chiral kinks depending on $|\theta|-\theta_0$ as in  \cite{AlyoshaTBA}:
\vskip .1cm
\hskip 2cm\includegraphics[height=6cm]{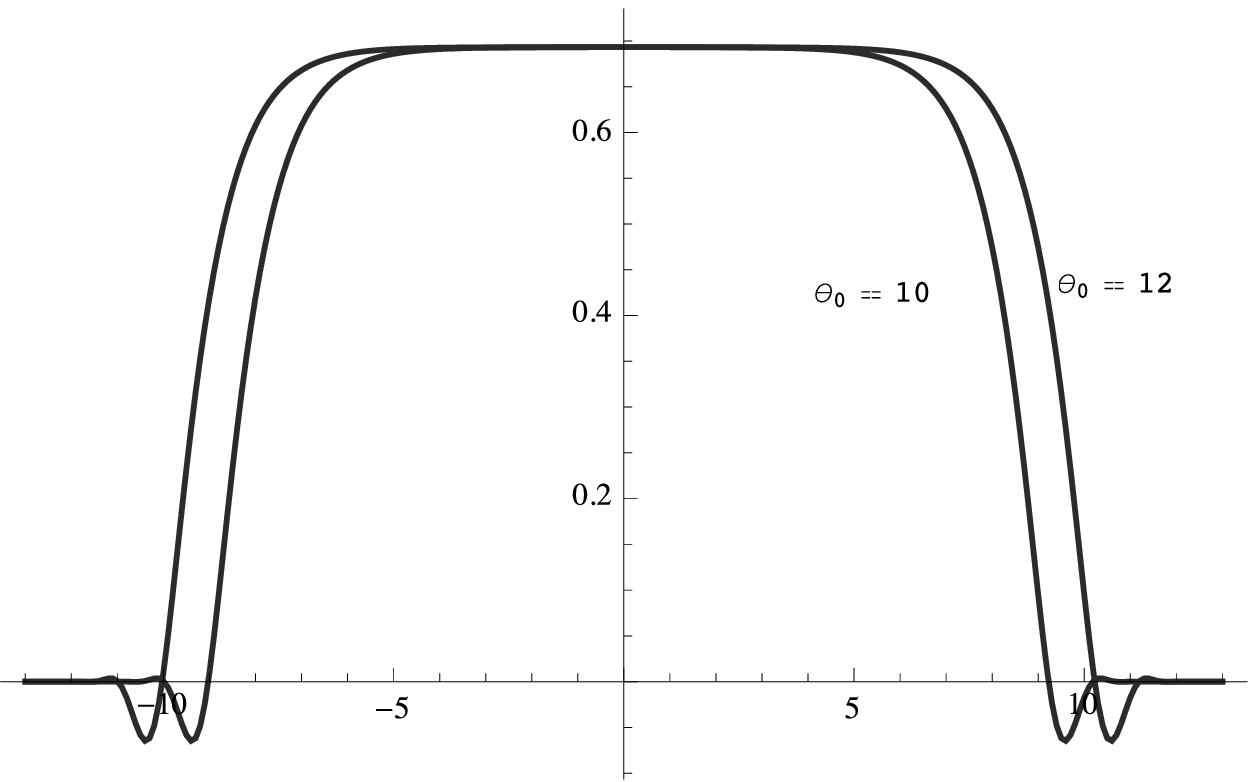}
\newline
{\it fig.1a }  Real part of $\log\(1+\mathfrak{a}\(\theta+\pi i\eta\)\)$ for $\theta_0=10,12$.
\vskip .2cm
\vskip .1cm
\hskip 2cm\includegraphics[height=6cm]{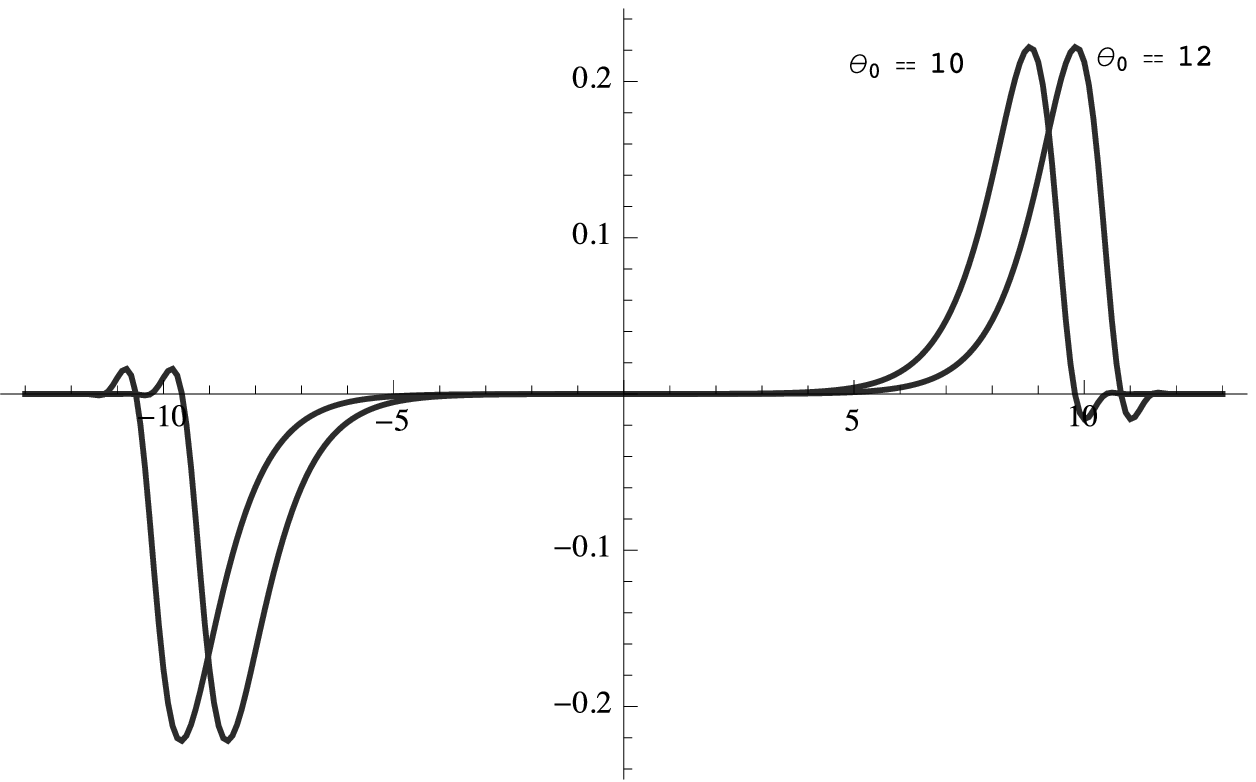}
\newline
{\it fig.1b } Imaginary part of $\log\(1+\mathfrak{a}\(\theta+\pi i \eta\)\)$ for $\theta_0=10,12$.
\vskip .2cm

The equations \eqref{DDV} are written in an assumption that for the ground state
all the Bethe roots are real. This is definitely true for $p=0$ and for sufficiently small
deviations from this point. Starting from some value of $p$ a rearrangement
of the Bethe roots begins. Experience shows that for $\nu$ close to the free fermion 
point $\nu =1/2$ we can go up to $p=0.2$ with equations \eqref{DDV}, for $\nu=3/4$ 
the border drops to $p\approx 0.12$. It should be said also that with our procedure
the number of iterations needed in order to reach the precision of $10^{-12}$ grows
fast for $\nu>3/4$, so, we  never work above this point.

If $-\eta<\mathrm{\theta}<\eta $ we have
\begin{align}
\log\Rho(\theta|\theta_0,p,p')=\frac 1 {2\pi i}\int_{-\infty}^{\infty}\Bigl(\frac {L(\theta'+i\eta|\theta_0,p,p')} {\sinh(\theta-\theta'-i\eta)}-\frac {\overline{L(\theta'+i\eta|\theta_0, p,p')}} {\sinh(\theta-\theta'+i\eta)}
\Bigr)d\theta'\,,\label{logr1}
\end{align}
where
$$L(\theta|\theta_0,p,p')=\log\(1+\mathfrak{a}(\theta|\theta_0,p')\)-\log\(1+\mathfrak{a}(\theta|\theta_0,p)\)
\,.$$
Since $L(\theta,\eta|\theta_0,p,p')$ decreases at $\theta\to\infty$ faster that any power of $e^{-\theta}$ we obtain the asymptotic expansion for $\theta\to\pm \infty$ in terms of $e^{\mp(2j-1)\theta}$. This gives the eigenvalues of 
local integrals of motion
\begin{align}
&\log\Rho(\theta|\theta_0,p,p')\ \ \simeq\hskip -.8cm\raisebox{-.3cm}{\scalebox{.75}{${\theta\to\infty}$}}\ \ \sum_{j=1}^{\infty}e^{-(2j-1)(\theta-\theta_0)}C_{2j-1}(\nu)\Bigl(I_{2j-1}(p',\theta_0)-I_{2j-1}(p,\theta_0)\Bigr)\label{asrho}\,,\\
&I_{2j-1}(p,\theta_0)=\frac {2e^{-(2j-1)\theta _0}} {\pi C_{2j-1}(\nu)}\int_{-\infty}^{\infty}\mathrm{Im}\Bigl(e^{(2j-1)(\theta+i\eta)}
\log\(1+\mathfrak{a}(\theta+i\eta|\theta_0,p)\)
\Bigr)d\theta\,.\nn
\end{align}
and similarly for the second chirality.
With this we can check that for  high temperature limit 
\begin{align}
I_{2j-1}(p,\theta_0)\ \ \to\hskip -.8cm\raisebox{-.3cm}{\scalebox{.75}{${\theta_0\to\infty}$}}\ \ I_{2j-1}(p)\,.\nn
\end{align}
Numerically for $\theta_0=11$, $p=0.15$, $\nu=5/8$ we arrive at precision of ten significant digits after 23
iterations of DDV equation for the first  three integrals \eqref{eigenI}. We do not go into details
because we use this well-known \cite{BLZ} computation just for verification  of
our numerics.

We shall call the asymptotical expansion in $e^{-(2j-1)(|\theta|-\theta_0)}$ of the kind of \eqref{asrho} the 
{\it standard} asymptotics in $\theta$.
The natural temptation is to consider $|\theta|-\theta_0$ as large as possible. 
However, this is dangerous because the absolute values of numerical answers would be too
small. We have to optimise somehow. Experience shows that for the computations below the
good choice is with $|\theta|-\theta_0 $ between $2$ and $3$, for this values  $e^{-2(|\theta|-\theta_0)}$ is
between $0.02$ and $0.0025$, which happens to be a reliable asymptotical domain.

In what follows we shall need in addition to $\Rho(\theta|\theta_0)$ for real $\theta$, the same function
for $\mathrm{Im}(\theta)=\eta$. In principle this can
be computed going to the border of the strip in  \eqref{logr1} via Sokhotsky's formula. However, it
is hard to use the latter  numerically with good precision. So, as has been mentioned, we solve 
the  equations \eqref{DDV} for the shift equal $\eta/2$ and proceed by
\begin{align}
\log\Rho(\theta|\theta_0,p,p')&=\pm L(\theta|\theta_0,p,p')\label{logr2}\\&+\frac 1 {2\pi i}\int_{-\infty}^{\infty}\Bigl(\frac {L(\theta'+i\eta/2|\theta_0,p,p')} {\sinh(\theta-\theta'-i\eta/2)}-\frac {\overline{L(\theta'+i\eta/2|\theta_0, p,p')}} {\sinh(\theta-\theta'+i\eta/2)}
\Bigr)d\theta'\,,\nn
\end{align}
which is valid for $\mathrm{Im}(\theta)\ \raisebox{.1cm}{$>$}\hskip -.3cm \raisebox{-.1cm}{$<$}\hskip .2cm\pm\eta/2$. 

To compute the asymptotics for any argument we use \eqref{logr1} because the analytical continuation
add a function decreasing faster that any $e^{N|\theta|}$ for any $N$ (see \eqref{logr2}).

{To make the formulae more readable, from now on
 we shall put the parameters $\theta_0,p,p'$ explicitly only when it is really needed. }

Now we formulate appropriate for numerical study
equations for the function
$$\Omega(\sigma-{\textstyle \frac{\pi i}2},\tau-{\textstyle \frac{\pi i}2}|\al)\,,$$
the arguments are shifted for convenience. 
Using this function
one can compute the expectation values of local operators on the infinite cylinder with
the boundary conditions with twists $p,p'$, as explained in the previous section. We follow the Appendix of 
\cite{HGSIV} with somewhat different notations and application. First, we switch to rapidity-like
variables. Second, we deal not with chiral CFT, but with sG, but this does not change much
for derivation of equations. Finally, for the auxiliary kernel $\psi$ we take
$$\psi(\theta|\al)=\frac{e^{\al\nu\theta}}{e^{2\nu\theta}-1}\,.$$

Introduce the kernel
$$K(\theta,\theta '|\al)=\frac {\nu} {\pi i}\Bigl(\psi(\theta-\theta'+\pi i|\al)
-\psi(\theta-\theta'-\pi i|\al)
\Bigr)\frac 1{\Rho(\theta')}\,,$$
and the corresponding integral operator
$$( {K}f)(\theta)=\int_{-\infty}^{\infty}K(\theta,\theta'|\al)f(\theta'){d\theta'}\,.$$
We shall need the resolvent $ {R}$ of $1- {K}$ satisfying
$$ {R}- {K} {R}= {K}\,,$$
and  solution to the equation
\begin{align}
&F_{\mathrm{right}}(\theta,\tau|\al)-
\int_{-\infty}^{\infty}K(\theta,\theta'|\al)F_{\mathrm{right}}(\theta',\tau|\al)d\theta'\label{eqFright}\\&=
\psi(\theta-\tau-i{\textstyle{\frac \pi 2}}|\al)\Rho^{-1/2}(\tau-i{\textstyle{\frac \pi 2}})-\psi(\theta-\tau+i{\textstyle{\frac \pi 2}}|\al)\Rho^{1/2}(\tau-i{\textstyle{\frac \pi 2}})\,.\nn
\end{align}
Here $\tau$ is an observer. 
Due to \cite{HGSIV} we know that $F_{\mathrm{right}}(\theta,\tau|\al)$ allows standard
asymptotics in  $\tau$. Let us concentrate on \eqref{eqFright}, the resolvent is treated similarly. 
Since $\Rho(\theta)\to 1$ at $\theta\to\pm\infty$ it does not make much sense to iterate \eqref{eqFright}:
the operator $ {K}$ is not compact. As usual we have to extract something in order to
reduce the operator to a compact one. Dividing the unknown function as
\begin{align}F_{\mathrm{right}}(\theta,\tau|\al)=F^{(0)}_{\mathrm{right}}(\theta,\tau)+F^\mathrm{corr}_{\mathrm{right}}(\theta,\tau|\al)\,,\quad
F^{(0)}_{\mathrm{right}}(\theta,\tau)=\frac{i}{2\nu}\frac {\Rho^{1/2}(\theta)}{\cosh(\theta-\tau)}\,,\label{defFcorr}\end{align}
and inverting an operator with difference kernel we arrive at the equation
\begin{align}
&F^\mathrm{corr}_{\mathrm{right}}(\theta,\tau|\al)-\int_{-\infty}^{\infty}G(\theta-\theta'|\al)\Bigl(\frac 1{\Rho(\theta')}-1\Bigr)F^\mathrm{corr}_{\mathrm{right}}(\theta',\tau|\al)
d\theta'\label{eqFcorr}\\&
=\Psi_-(\theta-\tau|\al)\Rho^{-1/2}(\tau-i{\textstyle{\frac \pi 2}})
-\Psi_+(\theta-\tau|\al)\Rho^{1/2}(\tau-i{\textstyle{\frac \pi 2}})
\nn\\&-F^{(0)}_{\mathrm{right}}(\theta,\tau)+\int_{-\infty}^{\infty}G(\theta-\theta'|\al)\Bigl(\frac 1{\Rho(\theta')}-1\Bigr)F^{(0)}_{\mathrm{right}}(\theta',\tau)
d\theta'\,,\nn
\end{align}
where
\begin{align}
&G(\theta|\al)=\int_{-\infty}^{\infty}e^{i\theta x}\frac{\sinh\frac\pi 2\(\frac{2\nu-1}{\nu}x-i\al\)}
{4\nu\sinh\frac\pi 2\(\frac{1-\nu}{\nu}x+i\al\)\cosh\(\frac{\pi x} 2\)}dx\,,\nn\\
&\Psi_{\pm}(\theta|\al)=\int_{-\infty}^{\infty}e^{i\theta x}\frac{\exp\(\pm\frac\pi 2\(\frac{1-\nu}{\nu}x+i\al\)\)}
{8i\nu\sinh\frac\pi 2\(\frac{1-\nu}{\nu}x+i\al\)\cosh\(\frac{\pi x} 2\)}dx\,.\nn
\end{align}
Consider $\tau>0$. 
It is easy to see that the right hand side of \eqref{eqFcorr} has standard asymptotics. In order
to find the coefficients of the asymptotic expansion 
we compute the right hand side at
five points  $\tau=\theta_0+2+2j/10$, $j=0,\cdots,4$ and use the interpolation procedure which
defines the first coefficient of the asymptotical series with precision $O(e^{-20})$ and the
second coefficient with the precision $O(e^{-16})$. Certainly, we  hope that there are no
resonances in higher terms which happens to be the case since our procedure works 
perfectly well. 
We work at given numerical precision, 
and taking more interpolation points would be not only useless, but even dangerous.
Then we solve the equation \eqref{eqFcorr} putting these coefficients into 
right hand side. 
As a result we obtain
$$F^\mathrm{corr}_{\mathrm{right}}(\theta,\tau|\al,\theta_0)=F^\mathrm{corr}_{\mathrm{right},1}(\theta,\theta_0|\al)e^{-(\tau-\theta_0)}+F^\mathrm{corr}_{\mathrm{right},3}(\theta,\theta_0|\al)e^{-3(\tau-\theta_0)}+\cdots\,.$$
Consistence with the high temperature behaviour exhibited on {\it fig.1a, fig.1b} requires that 
$F^\mathrm{corr}_{\mathrm{right},2j-1}(\theta,\theta_0|\al)$ depends on $\theta-\theta_0$.
Our numerics supports this as demonstrated on {\it fig.2} below,

 \vskip .1cm
\hskip 2cm\includegraphics[height=6cm]{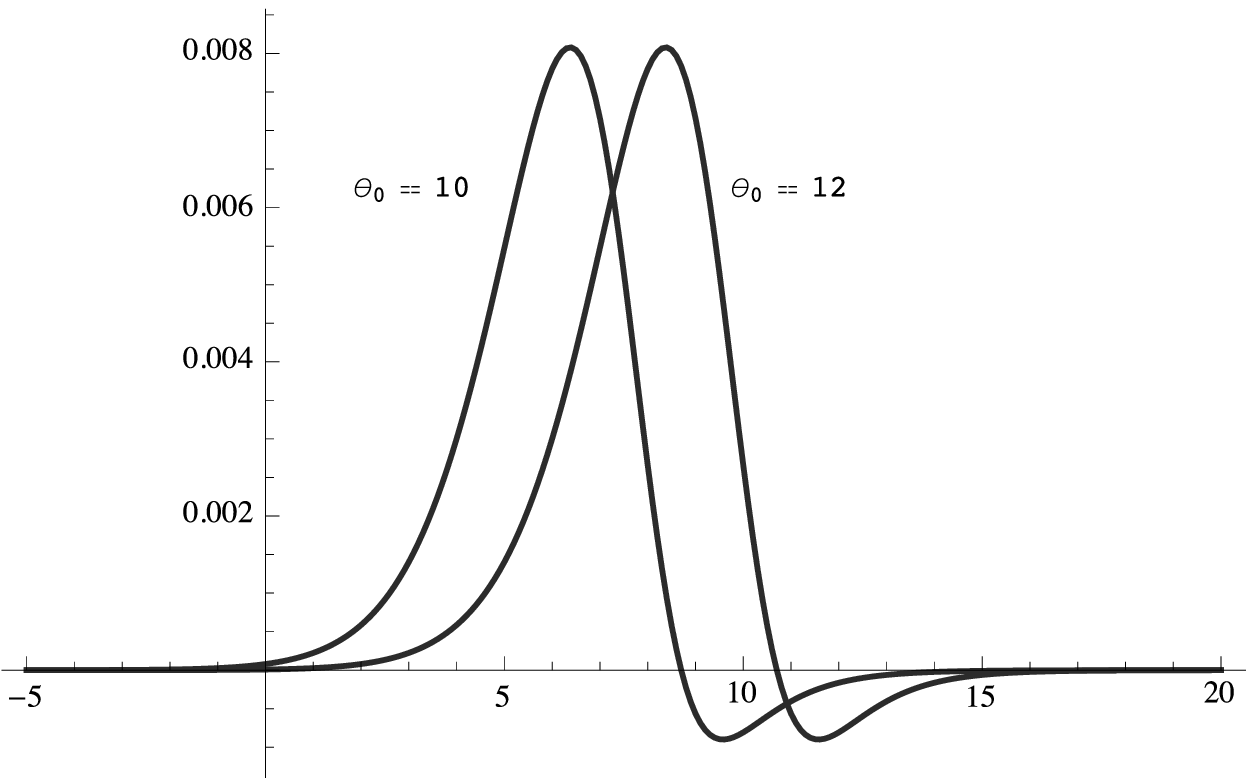}
\newline
{\it fig.2} {$iF^\mathrm{corr}_{\mathrm{right},1}(\theta,\theta_0|\al)$} for $\al=5/7,\nu=5/8$ and $\theta_0=10,12$.

\vskip.2cm



We work on the finite interval $[-\theta_\mathrm{max},\theta_\mathrm{max}]$ choosing $\theta_\mathrm{max}$ from the requirement that at the ends  the function does not
exceed $10^{-12}$. For example, for $\al=5/7,\nu=5/8, \theta_0=11$ it is sufficient to take $\theta_\mathrm{max}=35$.
For small $\al$ the necessary value of $\theta_\mathrm{max}$ grows which is not surprising 
for $\Omega(\sigma,\tau|\al)$ has a pole at $\al=0$. That is why we avoid  values of $\al$ smaller than $1/3$.

We shall need also
$$F_{\mathrm{left}}(\sigma,\theta|\al)=F_{\mathrm{right}}(\theta,\sigma|2-\al)\,.$$

Now we are ready to  define $\Omega(\sigma-\textstyle \frac{\pi i}2, \tau-\textstyle \frac{\pi i}2|\al)$. It consists of two parts:
$$\Omega(\sigma-{\textstyle \frac{\pi i}2}, \tau-{\textstyle \frac{\pi i}2}|\al)=\Omega^{(1)}(\sigma, \tau|\al)+\Omega^{(2)}(\sigma, \tau|\al)\,.
$$
The second part is rather explicit, it will be discussed later. In order to describe  the first part we 
introduce the measure
\begin{align}M^-(\theta)=\frac 1 {\Rho(\theta-i\eta|\theta_0,p,p')(1+\mathfrak{a}(\theta-i\eta|\theta_0,p))}\,,\quad
M^+(\theta)=\overline{M^-(\theta)}\,,\label{measure}\end{align}
the notations
$$F^\epsilon_{\mathrm{right}\atop\mathrm{left}}(\theta,\tau|\al)=F_{\mathrm{right}\atop\mathrm{left}}(\theta+ i\epsilon\eta,\tau|\al)\,,\quad
R^{\epsilon,\epsilon'}(\theta,
\theta'|\al)=R(\theta+i\epsilon \eta,\theta'+i\epsilon' \eta|\al)\,,\quad \epsilon,\epsilon'=\pm\,,$$
and a system of linear integral equations for two functions $G^\pm(\theta,\tau|\al)$:
\begin{align}
G^{\epsilon}(\theta,\tau|\al)=F^{\epsilon}_\mathrm{right}(\theta,\tau|\al)&-\sum\limits_{\epsilon'=\pm}
\ \ \int\limits_{-\infty}^{\infty}R^{\epsilon,\epsilon'}(\theta,
\theta'|\al)G^{\epsilon'}(\theta',\tau|\al)M^{\epsilon'}(\theta')d\theta'\,,\ \ \epsilon=\pm\label{eqG}\,.
\end{align}
Actually $G^\pm(\theta,\tau|\al)=G(\theta\pm i\eta,\tau|\al)$ for certain analytical function $G(\theta,\tau|\al)$,
but this is irrelevant for actual computation which consists in iterating the equation \eqref{eqG}.
Iterations converge fast because of very rapid decrease of $M^{\pm}(\theta)$ at $\theta\to\pm\infty$.
Finally,
\begin{align}
\Omega^{(1)}(\sigma, \tau|\al)=\frac{\nu}{\pi i }
\sum\limits_{\epsilon=\pm}
\ \ \int\limits_{-\infty}^{\infty}F_\mathrm{left}^{\epsilon}(\sigma,
\theta|\al)G^{\epsilon}(\theta,\tau|\al)M^{\epsilon}(\theta)d\theta\,.\nn
\end{align}
It is easy to see that
$$\Omega^{(1)}(\sigma, \tau|2-\al)=\Omega^{(1)}(\tau,\sigma|\al)\,.
$$

In order to define $\Omega^{(2)}(\sigma, \tau|\al,\theta_0)$ we introduce the function
\begin{align}
U(\theta,\tau)&=\half \tanh\half(\theta-\tau)\Bigl(\Rho(\theta)^{\frac 1 2}\Rho(\tau)^{\frac 1 2}-\Rho(\theta)^{-\frac 1 2}\Rho(\tau)^{-\frac 1 2}\Bigr)\nn\\&+\half \coth\half(\theta-\tau)\Bigl(\Rho(\theta)^{\frac 1 2}\Rho(\tau)^{-\frac 1 2}-\Rho(\theta)^{-\frac 1 2}\Rho(\tau)^{\frac 1 2}\Bigr)\,.\nn
\end{align}
Then
\begin{align}
\Omega^{(2)}(\sigma, \tau|\al)=\frac 1 {4\nu}U(\sigma-{\textstyle \frac{\pi i} 2},\tau-{\textstyle \frac{\pi i} 2}){+\frac 1 {2\pi i}}
\int\limits_{-\infty}^\infty U(\sigma-{\textstyle \frac{\pi i} 2},\theta)
F_\mathrm{right}(\theta,\tau)\frac{d\theta}{\Rho(\theta)^{\frac 1 2 }}\,.\nn
\end{align}
It is less obvious, but still can be shown, that
$$\Omega^{(2)}(\sigma, \tau|2-\al)=\Omega^{(2)}(\tau,\sigma|\al)\,.
$$
Let us divide $\Omega^{(2)}(\sigma, \tau|\al)$ into two pieces according to 
\eqref{defFcorr}:
\begin{align}
&\Omega^{(2)}(\sigma, \tau|\al)=\Omega^{(3)}(\sigma, \tau|\al)+\Omega^{(4)}(\sigma, \tau)\,,\nn\\
&\Omega^{(3)}(\sigma, \tau|\al)={\frac 1 {2\pi i}}
\int\limits_{-\infty}^\infty U(\sigma-{\textstyle \frac{\pi i} 2},\theta)
F_\mathrm{right}^\mathrm{corr}(\theta,\tau)\frac{d\theta}{\Rho(\theta)^{\frac 1 2 }}\,,\nn\\
&\Omega^{(4)}(\sigma, \tau)=\frac 1 {4\nu}\Bigl(U(\sigma-{\textstyle \frac{\pi i} 2},\tau-{\textstyle \frac{\pi i} 2}){+\frac 1 {\pi}}
\int\limits_{-\infty}^\infty U(\sigma-{\textstyle \frac{\pi i} 2},\theta)
\frac 1
{\cosh(\theta-\tau)}d\theta\Bigr)\,.\nn
\end{align}

All of our functions $\Omega^{(1)}(\sigma, \tau|\al)$, $\Omega^{(3)}(\sigma, \tau|\al)$, $\Omega^{(4)}(\sigma, \tau)$
have standard asymptotics in both arguments. Let us consider them one by one.

For $\Omega^{(1)}(\sigma, \tau|\al)$ this follows from the
standard asymptotics in $\sigma, \tau$ of $F_\mathrm{left}(\sigma, \theta|\al)$, $F_\mathrm{right }(\theta,\tau|\al)$.
These  asymptotics contain growing in $\theta$ terms because we have to treat 
$F^{(0)}_\mathrm{left}(\sigma, \theta)$, $F^{(0)}_\mathrm{right }(\theta,\tau)$ as geometrical progressions.
However, this is harmless 
since $M^\epsilon(\theta)$
decreases faster than any exponent $e^{-N|\theta|}$.

For $\Omega^{(3)}(\sigma, \tau|\al)$
we use coefficients of the standard asymptotics  of $F^\mathrm{corr}_\mathrm{right }(\theta,\tau|\al)$, compute
for five values of $\sigma$ and apply 
the five point interpolation as has been discussed. The precision is very good.

Finally, for the independent of $\al$ piece $\Omega^{(4)}(\sigma, \tau)$ the 
standard asymptotics is proved by more 
delicate means, we refer to \cite{HGSIV} for explanation. 
We compute at five points of both $\sigma $ and  $\tau$ (25 points all together), and then apply the five point
interpolation for both arguments. 
Numerically this is the most complicated piece, 
a lot of precision is needed. 
On the other hand, looking more attentively at this function we observe that it changes sign under $p\leftrightarrow p'$, while the entire function $\Omega$, as has
been said,  is supposed
to be symmetric. This observation results in the following final procedure.

We compute the coefficients of the standard asymptotics in both arguments for 
\begin{align}\Omega^{(1)}(\sigma, \tau|\al)+\Omega^{(2)}(\sigma, \tau|\al)\,,\label{1+2}\end{align}
for $p,p'$. Then we do the same for the opposite order $p',p$, and take symmetric and
anti-symmetric with respect to $p\leftrightarrow p'$ parts. We make sure that the anti-symmetric
part does not depend on $\al$, this fact is observed with very good numerical precision. 
Then we check that this anti-symmetric part cancels with  $\Omega^{(4)}(\sigma, \tau)$.
This is demonstrated with very good numerical
precision for the first in both variables term
of asymptotics; for the second term the
precision is reasonably good (like five significant digits), it is possible to 
to improve the precision of $\Omega^{(4)}(\sigma, \tau)$, but from the present discussion it
follows that this is not needed: it is much more precise to take the symmetric part
of \eqref{1+2} as the final answer. For the latter we change the precision of computations
(taking step $1/20$ instead of $1/10$, computing on longer intervals {\it etc.}) and observe
that this does not change the answer significantly. 

The entire procedure is complicated but doable. Let us discuss the results. 

First, we compare the ``theoretical" prediction for  $\Omega_{1,1}(\al|\theta_0,p,p') $ with the numerical results for 
$\theta_0=11, \al=5/7, \nu=5/8$ and different $p,p'$.

\vskip.3cm

\centerline{
\begin{tabular}{|l|c|r|}
  \hline
  \ &\ &\ \\
  values of $p$, $p'$ & $e^{-2\theta_0}\Omega_{1,1}$ theoretical & $e^{-2\theta_0}\Omega_{1,1}$ numerical \\
  \ &\ &\ \\
  \hline
  $0$, $\ \ \ \,\,\ 0.15$ &$0.247951585\cdot 10^{-3} i$  & $0.247951583 \cdot 10^{-3} i$ \\ \hline
  $0.05$, $\ 0,18$  & $0.155706252 \cdot 10^{-2} i$ & $0.155706251\cdot 10^{-2} i$ \\ \hline
   $0.07$, $\ 0,12$  & $0.320971379\cdot 10^{-3}i$  & $0.320971379\cdot 10^{-3}i$  \\
  \hline
\end{tabular}
}

\vskip.3cm

\noindent
We see that the precision is very good. 

Now we do the same for  $\Omega_{1,3}(\al|\theta_0,p,p')  $ (still $\theta_0=11, \al=5/7, \nu=5/8$).
\vskip.3cm

\centerline{\begin{tabular}{|l|c|r|}
  \hline
  \ &\ &\ \\
  values of $p$, $p'$ & $e^{-4\theta_0}\Omega_{1,3}$ theoretical & $e^{-4\theta_0}\Omega_{1,3}$ numerical\\
  \ &\ &\ \\
  \hline
  $0$, $\ \ \ \,\,\ 0.15$ &$0.172386742\cdot 10^{-4} i$  & $0.172386746 \cdot 10^{-4} i$ \\ \hline
  $0.05$, $\ 0,18$  & $0.452152545 \cdot 10^{-4} i$ & $0.452152550\cdot 10^{-4} i$ \\ \hline
   $0.07$, $\ 0,12$  & $0.206735353\cdot 10^{-4}i$  & $0.206735354\cdot 10^{-4}i$  \\
  \hline
\end{tabular}}
\vskip.3cm

 \noindent
The precision is somewhat worse than in the previous case, but one has to have in mind that
some precision is lost when 
extracting of the second term of the asymptotics .

It is interesting to consider complex $\al$. The imaginary part should not be too big,
otherwise the number of iterations in the equations for $F_{\mathrm{right}}^\mathrm{corr}$ start
to grow fast. 
For $\theta_0=11,\al=5/7+i,\nu=5/8$ we have:
 
\vskip.3cm
\centerline{\begin{tabular}{|l|c|r|}
  \hline
  \ &\ &\ \\
 $2j-1,2k-1$ & $e^{-2(j+k-1)\theta_0}\Omega_{2j-1,2k-1}$ theoretical & $e^{-2(j+k-1)\theta_0}\Omega_{2j-1,2k-1}$ numerical \\
  \ &\ &\ \\
  \hline
  $1,1$ &$\begin{aligned} (&-0.231024785 \nn\\&+ 0.369863518 i)10^{-2}\nn\end{aligned} $ & 
  $\begin{aligned}(&-0.231024785 \nn\\&+ 0.369863517 i)10^{-2}\nn\end{aligned}$ \\ 
  \hline
  $1,3$  &$\begin{aligned}&-(0.871712558\\
 & + 0.433997337 i)10^{-4}\end{aligned} $ & $
  \begin{aligned}&-(0.871712559\\& + 0.433997353 i)10^{-4}\end{aligned} $ \\
  \hline
\end{tabular}}

\vskip.3cm

\noindent
The precision is quite good.

Now we consider the shift of the primary field comparing with 
the equation \eqref{3point}. Here the dependance on $\theta_0$
is non-trivial, so, the most interesting check consists in taking
different values of $\theta_0$ for the same $p,p',\al,\nu$.
The right hand side of \eqref{3point} consists of three parts, 
it is interesting to consider the case when all of them contribute significantly,
for example, we set for $p=.12,p'=.15, \al=3/2, \nu=3/5$.
 
\vskip.3cm

\centerline{\begin{tabular}{|l|c|r|}
  \hline
  \ &\ &\ \\
  values of $\theta_0$ & $10^4ie^{-2\theta_0}\Omega_{1,-1}$ theoretical & $10^4ie^{-2\theta_0}\Omega_{1,-1}$ numerical \\
  \ &\ &\ \\
  \hline
  10 & 0.6915132469    &  0.6915132506        \\ \hline
   10.2 &   0.6704669264   &     0.6704669284         \\ \hline
  10.4 &0.6517998575    & 0.6517998585             \\ \hline
 10.6 &  0.6352432119    &     0.6352432124      \\ \hline
  10.8 &  0.6205584892     &        0.6205584894      \\ \hline
  11 &0.6075341106      &    0.6075341106         \\ \hline
  \end{tabular}}

\vskip.3cm
\noindent
We see that the agreement is very good, and it is getting better with $\theta_0$ growing.
This is not surprising because with $\theta_0$ growing we are going deeper into the 
high temperature domain.
We want to see whether the three parts in \eqref{3point} contribute considerably. 
For $\theta_0=11$ we have:
\begin{align}
&10^4i\cdot(I)=0.5053760392\,,\nn\\
&10^4i\cdot(II)=0.0000012457\,,\nn\\
&10^4i\cdot(III)=0.1021593171\,.\nn
\end{align}
The second part is smaller that other two (this is not a wonder since it contains $e^{-2\theta_0}$), but still
it contributes five orders more than the discrepancy between theoretical and  numerical values. 

To finish this section let us give an example of computation of the ratio of
one-point functions for arbitrary $\theta_0$. We compute
$$\Lambda(p,p',\al,\nu,\theta_0)=\log\(-\frac {\langle p|\Phi_{\al+2\frac{1-\nu}{\nu}}(0)|p'\rangle} {\langle p|\Phi_\al(0)|p'\rangle}\)\,,$$
using \eqref{sss}. For $\theta_0\to \infty$, $\Lambda(p,p',\theta_0,\al,\nu)$ reproduces the CFT
result:
$$\Lambda(p,p',\al,\nu,\theta_0)\to \log(-C(p,p',\al,\nu))+2\bigl(\Delta_{\al+2\frac{1-\nu}{\nu}}-\Delta_{\al}\bigr)\theta_0\,.$$
For $\theta_0\to -\infty$, $\Lambda(p,p',\theta_0,\al,\nu)$ produces the ratio of Lukyanov-Zamolodchikov one-point
functions, shifts $p,p'$ become irrelevant in this limit,
$$\Lambda(p,p',\al,\nu,\theta_0)\to \log(-C(\al,\nu))\,.$$
An example of numerics is provided on {\it fig. 3}.
\vskip .1cm
\hskip 2cm\includegraphics[height=6cm]{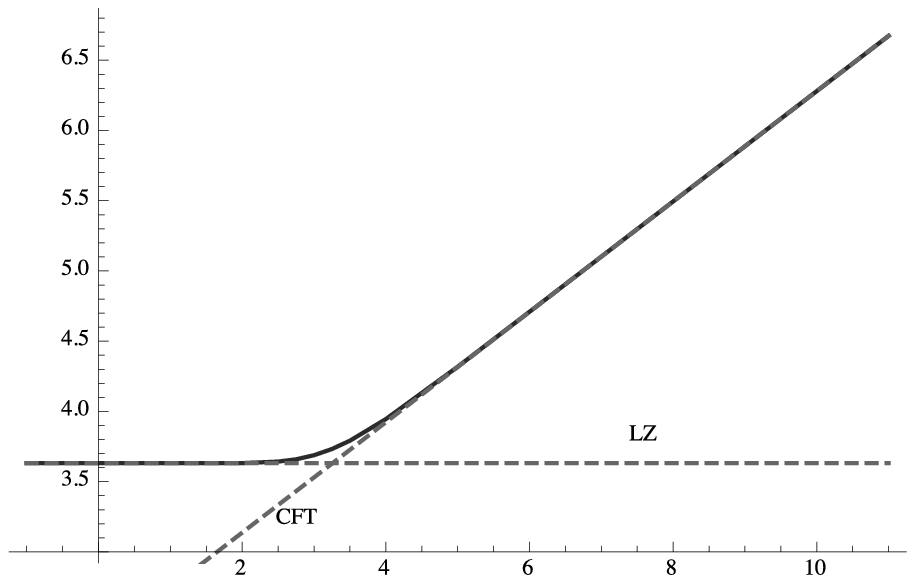}
\newline
{\it fig.3} The function $\Lambda(0,0.15,5/7,5/8,\theta_0)$ against high (CFT)  and low  (LZ) temperature predictions.
\vskip .1 cm

We observe very fast crossover. 
\section{Conclusion}

As before \cite{HGSV}, we obtain the one-point function not for the primary
fields, but for the ratio of primary fields shifted by $2\frac{1-\nu}{\nu}$. This is
sufficient for computing  the UV asymptotics of the normalised 
two-point function 
$$G_{\al_1,\al_2}(z,\bar z)=\frac{\langle\Phi_{\al_1}(z,\bar z)\Phi_{\al_2}(0)\rangle}{\langle\Phi_{\al}(0)\rangle}\,,\quad \al=\al_1+\al_2\,,$$
on a cylinder with different boundary conditions. 
Indeed, the PCFT OPE contains only the 
primary fields (and their descendants) shifted from $\Phi_{\al_1+\al_2}(0)$ by multiple of
$2\frac{1-\nu}{\nu}$. 
The PCFT contains rather complicated integrals which should be defined by analytical continuation.
In the domain $1/2<\nu<1$ and for positive $\al_1$, $\al_2$ in order to obtain the asymptotics of
$(z\bar z)^{\Delta_{\al_1}+\Delta_{\al_2}-\Delta_\al}G_{\al_1,\al_2}(z,\bar z)$ up to $o((z\bar z)^2)$
one needs only one of these integrals (which is known \cite{dotsenko}) and 
 the normalised by $\langle \Phi_\al\rangle$
one-point functions of $\Phi_{\al+2\frac{1-\nu}{\nu}}$,
$\mathbf{l}_{-2}\Phi_{\al}$, $\bar{\mathbf{l}}_{-2}\Phi_{\al}$, $\mathbf{l}_{-4}\Phi_{\al}$, $\bar{\mathbf{l}}_{-4}\Phi_{\al}$
and  $\mathbf{l}_{-2}\bar{\mathbf{l}}_{-2}\Phi_{\al}$
which we can compute with the results of this paper. The latter one-point function
is given by $2\times 2$ determinant which in the
simplest case $p=p'$, $\al=0$ coincides with Zamolodchikov's one \cite{Sasha} as explained in \cite{HGSV}.
We hope to return to computation of this asymptotics in future. 

Still knowledge of the one-point function itself is interesting.
Certainly, assuming analyticity we can just solve the difference equations for
$\langle\Phi_{\al}(0)\rangle$, and eliminate possible quasi-constant by reasonable
analyticity requirements obtaining the $\al$-dependent part of $\langle\Phi_{\al}(0)\rangle$,
but the problem of normalisation arises. In the case of
coinciding boundary condition this could be fixed normalising by the expectation value of $1$
(partition function).
In general case this is impossible because the expectation value of $1$ vanishes. 

The  above mentioned problem is even more serious if we try  to apply our formulae to compute
the form factor in finite volume: we obtain not the form factors, but their ratios.
So, the form factors of the primary fields should be defined. In this case
the normalisation depends on the quantum numbers of two states involved,
and it can be quite non-trivial. This happens even in the case of infinite volume,
as can be seen from the computations of \cite{JMSFF}. This deserves further 
investigation.

In this paper we provided a conclusive evidence for the construction of
the fermionic basis in the case when the local integrals of
motion act non-trivially. 
We used two main tools: the conjecture of the existence of the fermionic basis and
the reflection relations. 
This was done up to the level 4. Proceeding to higher levels would encounter 
certain technical problems. 

First, the fermionic basis modulo integrals of motion 
contain certain unusual in CFT denominators (the simplest one is $\Delta_\al+2$ on level 6).
This would require rethinking our minimality conditions. Second, the numerical data are less
precise, one would need to improve the working precision.
Since we do not know how to change the minimality conditions it would be desirable to
have enough numerical data for computing the coefficients $A$, $B$ numerically.
This is how we actually proceeded on levels 2 and 4. The coefficients in question
are obtained as follows. 

On level $4$ we have $5$ states in the Verma module which
we organise in our way: acting by even Virasoro generators and then by integrals of motion.
The normalised expectation values are given by five linearly independent polynomials of $p,p'$.
Correspondingly, we conjecture that the coefficients of the asymptotics are polynomials in $p,p'$.
So, we compute  numerically  asymptotics for different values of $p,p'$,
and, assuming that they are polynomials, present them as a linear combination of our five polynomials.
We compute for more than five pairs $p,p'$ in order to confirm our assumption.

On the level 6 there are eleven descendants, and only nine of corresponding polynomials are
linearly independent. The situation is similar to the one considered in \cite{Boos}:
to have enough data it is not sufficient to consider only the ground states of Matsubara transfer-matrices
as boundary conditions. One have to include at least one exited state.
This is a hard problem,
we hope to return to it in future.

In this paper we considered only chiral null-vectors which have well-known
counterpart in CFT. However, it is easy to generalise our procedure in order
to consider the null-vectors mixing two chiralities. They are nothing but higher equations
of motion (compare with the Liouville case \cite{AlyoshaHigher})
the sG equation itself being the simplest example. The derivatives of the field $\varphi(z,\bar z)$
occur from the limit of $t_0(\al)\mathbf{i}_{2j-1}\Phi_\al$ for $\al\to 0$. 

\section{Appendix}
Here we give explicit formulae for functions $X_{\#}^{\#}(\al,\nu)$:
\begin{align}
       &X_{1,3}^{1,3}(\al,\nu)=   -\frac 4 {27 \al ( (\al \nu)^2-4)   
      ( (\al \nu)^2-4(1-\nu)^2) }  \nn\\
      &  (36 + 141 \al - 72 \nu - 342 \al \nu + 60 \al^2 \nu + 36 \nu^2 + 
       291 \al \nu^2 - 186 \al^2 \nu^2 - 16 \al^3 \nu^2\nn\\& - 90 \al \nu^3 + 
       174 \al^2 \nu^3 - 16 \al^3 \nu^3 - 16 \al^4 \nu^3 - 72 \al^2 \nu^4 + 
       8 \al^4 \nu^4)\,,\nn\\ \ \nn\end{align}\begin{align}
     &  X_{1,3}^{1,1|1,1}(\al,\nu)=\frac 1 {9 (  \nu-1) ( (\al \nu)^2-4)   
      ( (\al \nu)^2-4(1-\nu)^2) (-1 + \nu + \al \nu)}\nn\\
       &2 (-84 + 384 \nu + 36 \al \nu - 720 \nu^2 - 24 \al \nu^2 + 
       38 \al^2 \nu^2 + 696 \nu^3 \nn\\&- 168 \al \nu^3 - 158 \al^2 \nu^3 + 
       18 \al^3 \nu^3 - 348 \nu^4 + 300 \al \nu^4 + 226 \al^2 \nu^4 - 
       69 \al^3 \nu^4 \nn\\&- 4 \al^4 \nu^4 + 72 \nu^5 - 180 \al \nu^5 - 
       118 \al^2 \nu^5 + 60 \al^3 \nu^5 - 3 \al^4 \nu^5 + 36 \al \nu^6\nn\\& - 
       24 \al^2 \nu^6 + 3 \al^3 \nu^6 + 24 \al^4 \nu^6 - 3 \al^5 \nu^6 - 
       2 \al^6 \nu^6 + 36 \al^2 \nu^7 - 13 \al^4 \nu^7 + 
       \al^6 \nu^7)\,,\nn\\ \ \nn\end{align}\begin{align}
         &X_{1,3}^{1,1,1,1}(\al,\nu)    =\frac{1}{27 \al (\nu-1) ( (\al \nu)^2-4)   
      ( (\al \nu)^2-4(1-\nu)^2) } \nn\\&
              (720 + 660 \al - 2448 \nu - 2028 \al \nu + 336 \al^2 \nu + 3024 \nu^2 + 
     1872 \al \nu^2 - 1476 \al^2 \nu^2 \nn\\& + 29 \al^3 \nu^2 - 1584 \nu^3 - 
     60 \al \nu^3 + 2040 \al^2 \nu^3 - 342 \al^3 \nu^3 - 52 \al^4 \nu^3 + 
     288 \nu^4 - 804 \al \nu^4 \nn\\& - 1044 \al^2 \nu^4 + 443 \al^3 \nu^4 + 
     18 \al^4 \nu^4 - 10 \al^5 \nu^4 + 360 \al \nu^5 + 24 \al^2 \nu^5 - 
     50 \al^3 \nu^5  \nn\\&+ 104 \al^4 \nu^5 - 10 \al^5 \nu^5 - 10 \al^6 \nu^5 + 
     180 \al^2 \nu^6 - 65 \al^4 \nu^6 + 
     5 \al^6 \nu^6)\,.\nn
\end{align}
\begin{align}
&X_{3,1}^{1,3}(\al,\nu)=\frac 4 {27 \al( (\al \nu)^2-4)   
      ( (\al \nu)^2-4(1-\nu)^2)}\nn\\&(-36 - 141 \al + 72 \nu + 222 \al \nu + 60 \al^2 \nu - 36 \nu^2 - 
     111 \al \nu^2 + 6 \al^2 \nu^2 + 16 \al^3 \nu^2 \nn\\&+ 30 \al \nu^3 - 
     18 \al^2 \nu^3 - 48 \al^3 \nu^3 - 16 \al^4 \nu^3 + 24 \al^2 \nu^4 + 
     32 \al^3 \nu^4 + 8 \al^4 \nu^4)\,,\nn\\
     \ \nn\end{align}\begin{align}
&X_{3,1}^{1,1|1,1}(\al,\nu)=-\frac2 {9 (1 - \nu) (1 + \al \nu)((\al \nu)^2-4)   
      ( (\al \nu)^2-4(1-\nu)^2) }\nn\\&(-84 + 204 \nu - 36 \al \nu - 180 \nu^2 + 192 \al \nu^2 + 38 \al^2 \nu^2 + 
     84 \nu^3 - 252 \al \nu^3 - 32 \al^2 \nu^3\nn\\& - 18 \al^3 \nu^3 - 24 \nu^4 + 
     108 \al \nu^4 - 26 \al^2 \nu^4 + 3 \al^3 \nu^4 - 4 \al^4 \nu^4 - 
     12 \al \nu^5 + 8 \al^2 \nu^5 \nn\\&+ 39 \al^3 \nu^5 + 15 \al^4 \nu^5 - 
     24 \al^2 \nu^6 - 12 \al^3 \nu^6 + 6 \al^4 \nu^6 - 3 \al^5 \nu^6 - 
     2 \al^6 \nu^6 - 12 \al^3 \nu^7 \nn\\&- 4 \al^4 \nu^7 + 3 \al^5 \nu^7 + 
     \al^6 \nu^7)\,,\nn\end{align}\begin{align}
&X_{3,1}^{1,1,1,1}(\al,\nu)=-\frac 1 {27 \al (\nu-1)((\al \nu)^2-4)   
      ( (\al \nu)^2-4(1-\nu)^2) }\nn\\&(-720 - 660 \al + 1872 \nu + 1932 \al \nu + 336 \al^2 \nu - 1584 \nu^2 - 
   1632 \al \nu^2 - 204 \al^2 \nu^2\nn\\& - 29 \al^3 \nu^2 + 432 \nu^3 + 
   348 \al \nu^3 - 504 \al^2 \nu^3 - 226 \al^3 \nu^3 - 52 \al^4 \nu^3 + 
   132 \al \nu^4 \nn\\&+ 420 \al^2 \nu^4 + 409 \al^3 \nu^4 + 138 \al^4 \nu^4 + 
   10 \al^5 \nu^4 - 120 \al \nu^5 - 168 \al^2 \nu^5 - 74 \al^3 \nu^5\nn\\& - 
   16 \al^4 \nu^5 - 30 \al^5 \nu^5 - 10 \al^6 \nu^5 - 60 \al^2 \nu^6 - 
   80 \al^3 \nu^6 - 5 \al^4 \nu^6 + 20 \al^5 \nu^6 + 
   5 \al^6 \nu^6)\,.\nn
\end{align}

{\it Acknowledgements.}\quad FS is grateful to M. Jimbo and T. Miwa for numerous interesting
discussions of null-vectors in fermionic basis. 
HB would like to thank {\it Deutsche Forschungsgemeinschaft}
for the financial support of his project BO 3401/1--1 within the 
project {\it ``Forschergruppe''}.


\begin{thebibliography}{99}

\bibitem{HGS}
H.~Boos, M.~Jimbo, T.~Miwa, F.~Smirnov, and Y.~Takeyama.
\newblock Hidden {Grassmann} structure in the {XXZ} model.
\newblock {\em Commun. Math. Phys.} {\bf 272} (2007) 263--281. 

\bibitem{HGSII}
H.~Boos, M.~Jimbo, T.~Miwa, F.~Smirnov, and Y.~Takeyama.
\newblock Hidden {Grassmann} structure in the {XXZ} model {II : Creation}
  operators.
\newblock {\em Commun. Math. Phys.} {\bf 286} (2009) 875--932. 

\bibitem{HGSIII}
M.~Jimbo, T.~Miwa, and F.~Smirnov.
\newblock Hidden {Grassmann} structure in the {XXZ} model {III}: {Introducing
  Matsubara} direction.
\newblock {\em J. Phys. A} {\bf 42} (2009)  304018 (31pp)

\bibitem{HGSIV}
 H.Boos, M.~Jimbo, T.~Miwa, F. Smirnov,
\newblock Hidden {Grassmann} structure in the {XXZ} model {IV}: {CFT} limit.
\newblock {\em Commun. Math. Phys.}, {\bf 299} (2010) 825--866


\bibitem{HGSV}
 M.~Jimbo, T.~Miwa, F. Smirnov,
\newblock Hidden {Grassmann} structure in the {XXZ} model {V}: {sine-Gordon}
 model.
\newblock {\em Lett. Math. Phys.}, {\bf 96} (2011) 325--365

\bibitem{Alyosha2p}
Al. Zamolodchikov.
\newblock Two point correlation function in scaling {Lee-Yang} model.
\newblock {\em Nucl. Phys.} {\bf B348} (1991) 619--641. 



\bibitem{FFLZZ}
V.~Fateev, D.~Fradkin, S.~Lukyanov, A.~Zamolodchikov, and Al. Zamolodchikov.
\newblock Expectation values of descendent fields in the {sine-Gordon} model.
\newblock {\em Nucl. Phys.} {\bf B540} (1999) 587--609. 

\bibitem{Lukyanov} S. Lukyanov. Low energy effective Hamiltonian for the XXZ spin chain
{\em  Nucl.Phys.} {\bf B522} (1998) 533-549




\bibitem{OP}
M.~Jimbo, T.~Miwa, and F.~Smirnov.
\newblock On one-point functions of descendants in {sine-Gordon} model.
\newblock {\em New Trends in Quantum Integrable Systems}, World Scientific, (2009)
117-135

\bibitem{JMSFF}

M.~Jimbo, T.~Miwa, F. Smirnov, \newblock
Fermionic structure in the sine-Gordon model: Form factors and null-vectors
\newblock {\em Nuclear Physics B}
{\bf 852} (2011) 390-440


 \bibitem{book}
F. A. Smirnov
Form Factors in Completely Integrable Models of Quantum Field Theory,
{\it   Adv. Series in Math Physics 14}, World Scientific,
Singapore, 1992, 208 pp.



\bibitem{BLZII}
V.~Bazhanov, S.~Lukyanov, and A.~Zamolodchikov.
\newblock Integrable structure of conformal field theory {II}. {Q}-operator and
  {DDV} equation.
\newblock {\em Commun. Math. Phys.} {\bf 190} (1997) 247--278. 




\bibitem{Boos}
H.~Boos.
\newblock Fermionic basis in conformal field theory and thermodynamic {Bethe
  Ansatz} for excited states.
\newblock {\em SIGMA}, {\bf 7} (2011) 007, 36


\bibitem{NS} S. Negro, F. Smirnov. \newblock Reflection Relations and Fermionic Basis
{\it Lett. in Math. Phys.}
 {\bf 103} (2013)   1293Ð1311


\bibitem{LZ}
S.~Lukyanov and A.~Zamolodchikov.
\newblock Exact expectation values of local fields in quantum {sine-Gordon}
  model.
\newblock {\em Nucl.Phys.}, {\bf B493} (1997) 571--587







\bibitem{BBS}
O.~Babelon, D.~Bernard, and F.~Smirnov.
\newblock Null-vectors in integrable field theory.
\newblock {\em Commun. Math. Phys.}, {\bf 186} (1997) 601--648





\bibitem{ZZ}
A.~Zamolodchikov and Al. Zamolodchikov.
\newblock Structure constants and conformal bootstrap in {Liouville} field
  theory.
\newblock {\em Nucl.Phys.} {\bf B477}(1996) 577--605. 

\bibitem{BKP}
A.~Kl\"umper, M.~Batchelor, and P.~Pearce.
\newblock Central charges of the 6- and 19-vertex models with twisted boundary
  conditions.
\newblock {\em J.Phys. A: Math.Gen.}, {\bf 24}, 3111--3133, 1991.

\bibitem{DDV}
C.~Destri and H.J. de~Vega.
\newblock Unified approach to thermodynamic {Bethe Ansatz} and finite size
  corrections for lattice models and field theories.
\newblock {\em Nucl.Phys.}, {\bf B438}, 413--454, 1995.


\bibitem{Alyoshascale}
Al.B.~Zamolodchikov.
\newblock
Mass scale in the sine-Gordon model and its reductions,
\newblock{\em IJMPA} {\bf A10} No. 8 (1995) 1125-1150

\bibitem{AlyoshaTBA}
Al.B.~ Zamolodchikov.
\newblock
Thermodynamic Bethe ansatz in relativistic models: Scaling 3-state potts and Lee-Yang models,
\newblock{\em Nuclear Physics}
{\bf 342B} (1990) 695-720

 \bibitem{BLZ}
V.~Bazhanov, S.~Lukyanov, and A.~Zamolodchikov.
\newblock Integrable structure of conformal field theory, quantum {KdV} theory
  and thermodynamic {Bethe} ansatz,
\newblock {\em Commun. Math. Phys.}, {\bf 177}, 381--398, 1996

\bibitem{AlyoshaHigher}
Al.Zamolodchikov. Higher Equations of Motion in Liouville Field Theory,
{\em Int.J.Mod.Phys. } {\bf A19S2} (2004) 510-523

\bibitem{dotsenko} V.~Dotsenko, M.~Picco, P.~ Pujoi,  Renormalization group calculation of correlation functions for the 2D random bound Ising and Potts models. {\em Nucl. Phys.} {\bf B455} (1995) 701-723 



\bibitem{Sasha}
A. Zamolodchikov, Expectation value of composite field $T\overline{T}$ in two-dimensional quantum field theory, hep/th 0401146v1, January 2004


\end{thebibliography}
\end{document}